\documentclass[journal]{IEEEtran}
\usepackage{amsmath,amsfonts}
\usepackage{algorithmic}
\usepackage{algorithm}
\usepackage{array}
\usepackage[caption=false,font=normalsize,labelfont=sf,textfont=sf]{subfig}
\usepackage{textcomp}
\usepackage{stfloats}
\usepackage{url}
\usepackage{verbatim}
\usepackage{graphicx}
\usepackage{cite}
\usepackage[nolist]{acronym}
\usepackage{siunitx}
\usepackage{rotating}
\usepackage{tikz}
\usepackage{pgfplots}
\pgfplotsset{width=7cm,compat=1.3}
\usetikzlibrary{shapes,arrows,positioning,calc, matrix,spy,patterns,decorations.pathreplacing,calligraphy,colorbrewer,fit,intersections,decorations.text,spy}
\usepgfplotslibrary{groupplots}
\usepgfplotslibrary{colorbrewer} %
\usepackage{booktabs}
\usepackage{bbm}
\usepackage{bm}
\usepackage{float}
\usepackage{filecontents}

\definecolor{DarkGreen}{rgb}{0.0, 0.6, 0.0}

\definecolor{kit-green100}{rgb}{0,.59,.51}
\definecolor{kit-green70}{rgb}{.3,.71,.65}
\definecolor{kit-green50}{rgb}{.50,.79,.75}
\definecolor{kit-green30}{rgb}{.69,.87,.85}
\definecolor{kit-green15}{rgb}{.85,.93,.93}
\definecolor{KITgreen}{rgb}{0,.59,.51}

\definecolor{KITpalegreen}{RGB}{130,190,60}
\colorlet{kit-maigreen100}{KITpalegreen}
\colorlet{kit-maigreen70}{KITpalegreen!70}
\colorlet{kit-maigreen50}{KITpalegreen!50}
\colorlet{kit-maigreen30}{KITpalegreen!30}
\colorlet{kit-maigreen15}{KITpalegreen!15}

\definecolor{KITblue}{rgb}{.27,.39,.66}
\definecolor{kit-blue100}{rgb}{.27,.39,.67}
\definecolor{kit-blue70}{rgb}{.49,.57,.76}
\definecolor{kit-blue50}{rgb}{.64,.69,.83}
\definecolor{kit-blue30}{rgb}{.78,.82,.9}
\definecolor{kit-blue15}{rgb}{.89,.91,.95}

\definecolor{KITyellow}{rgb}{.98,.89,0}
\definecolor{kit-yellow100}{cmyk}{0,.05,1,0}
\definecolor{kit-yellow70}{cmyk}{0,.035,.7,0}
\definecolor{kit-yellow50}{cmyk}{0,.025,.5,0}
\definecolor{kit-yellow30}{cmyk}{0,.015,.3,0}
\definecolor{kit-yellow15}{cmyk}{0,.0075,.15,0}

\definecolor{KITorange}{rgb}{.87,.60,.10}
\definecolor{kit-orange100}{cmyk}{0,.45,1,0}
\definecolor{kit-orange70}{cmyk}{0,.315,.7,0}
\definecolor{kit-orange50}{cmyk}{0,.225,.5,0}
\definecolor{kit-orange30}{cmyk}{0,.135,.3,0}
\definecolor{kit-orange15}{cmyk}{0,.0675,.15,0}

\definecolor{KITred}{rgb}{.63,.13,.13}
\definecolor{kit-red100}{cmyk}{.25,1,1,0}
\definecolor{kit-red70}{cmyk}{.175,.7,.7,0}
\definecolor{kit-red50}{cmyk}{.125,.5,.5,0}
\definecolor{kit-red30}{cmyk}{.075,.3,.3,0}
\definecolor{kit-red15}{cmyk}{.0375,.15,.15,0}

\definecolor{KITpurple}{RGB}{160,0,120}
\colorlet{kit-purple100}{KITpurple}
\colorlet{kit-purple70}{KITpurple!70}
\colorlet{kit-purple50}{KITpurple!50}
\colorlet{kit-purple30}{KITpurple!30}
\colorlet{kit-purple15}{KITpurple!15}

\definecolor{KITcyanblue}{RGB}{80,170,230}
\colorlet{kit-cyanblue100}{KITcyanblue}
\colorlet{kit-cyanblue70}{KITcyanblue!70}
\colorlet{kit-cyanblue50}{KITcyanblue!50}
\colorlet{kit-cyanblue30}{KITcyanblue!30}
\colorlet{kit-cyanblue15}{KITcyanblue!15}

\definecolor{KITbraun}{RGB}{167,130,46}

\definecolor{cb-1}{HTML}{4477AA}
\definecolor{cb-2}{HTML}{EE6677}
\definecolor{cb-3}{HTML}{228833}
\definecolor{cb-4}{HTML}{CCBB44}
\definecolor{cb-5}{HTML}{66CCEE}
\definecolor{cb-6}{HTML}{AA3377}
\definecolor{cb-7}{HTML}{BBBBBB}

\pgfplotscreateplotcyclelist{cb list}{
	cb-1,cb-2,cb-3,cb-4,cb-5,cb-6,cb-7
}

\let\Re\relax
\let\Im\relax
\DeclareMathOperator{\Re}{Re}
\DeclareMathOperator{\Im}{Im}

\newcommand{\RV}[1]{\mathsf{#1}}

\newcommand*{\vect}[1]{\boldsymbol{#1}}

\newcommand*{\e}{\mathrm{e}}

\let\j\relax
\newcommand{\j}{\mathrm{j}}

\newcommand{\expecv}[2]{\mathbb{E}_{#1} \! \left\{#2\right\}}

\newcommand*{\PD}{P_{\text{D}}}
\newcommand*{\PFA}{P_{\text{FA}}}

\usepackage{xcolor}
\usepackage{amsmath}
\usepackage{amssymb}
\usetikzlibrary{calc}
\usetikzlibrary{positioning}
\usetikzlibrary{patterns}
\usetikzlibrary{decorations.pathreplacing,calligraphy}

\tikzset{
	lsblock/.style={rectangle, thick, draw, minimum width=1.2cm, minimum height=0.6cm, rounded corners=1.6mm, font=\small, align=center}
}

\tikzset{
	aeblock/.style={rectangle, thick, draw, minimum width=2.4cm, minimum height=0.7cm, rounded corners=1.6mm, align=center}
}
\tikzset{
	smlblock/.style={rectangle, thick, draw, minimum width=1.8cm, minimum height=0.6cm, rounded corners=1.2mm, align=center}
}
\tikzset{
	smsblock/.style={rectangle, thick, draw, minimum width=0.6cm, minimum height=0.4cm, rounded corners=1.0mm,inner sep=0.05cm, align=center},
    pasblock/.style={rectangle, thick, draw, minimum width=0.7cm, minimum height=0.7cm, rounded corners=1.6mm, align=center}
}

\tikzstyle{surround} = [rectangle, rounded corners, draw=KITred, inner sep=0.25cm, dashed, thick]
\tikzstyle{surround_AE} = [rectangle, rounded corners, draw=KITred, inner sep=0.13cm, dashed, thick]

\usetikzlibrary{external}
\begin{document}
\begin{acronym}[]
    \acro{6G}{sixth generation}
    \acro{AE}{autoencoder}
    \acro{AIR}{achievable information rate}
    \acro{APSK}{amplitude and phase-shift keying}
    \acro{ASK}{amplitude shift keying}
    \acro{AWGN}{additive white Gaussian noise}
    \acro{BMD}{bit-metric decoding}
    \acro{CA}{cell-averaging}
    \acro{CFAR}{constant false alarm rate}
    \acro{CP}{cyclic prefix}
    \acro{CSI}{channel state information}
    \acro{DRT}{deterministic-random trade-off}
    \acro{DM}{distribution matcher}
    \acro{EPI}{entropy power inequality}
    \acro{FEC}{forward error correction}
    \acro{FFT}{fast Fourier transform}
    \acro{GMI}{generalized mutual information}
    \acro{IFFT}{inverse fast Fourier transform}
    \acro{ISAC}{integrated sensing and communications}
    \acro{LLR}{log-likelihood ratio}
    \acro{LUT}{look-up table}
    \acro{MF}{matched filter}
    \acro{MI}{mutual information}
    \acro{OFDM}{orthogonal frequency division multiplexing}
    \acro{PAS}{probabilistic amplitude shaping}
    \acro{PDF}{probability density function}
    \acro{PSK}{phase shift keying}
    \acro{QPSK}{quadrature phase shift keying}
    \acro{RCS}{radar cross section}
    \acro{QAM}{quadrature amplitude modulation}
    \acro{RV}{random variable}
    \acro{SC}[S\&C]{sensing and communications}
    \acro{SMD}{symbol-metric decoding}
    \acro{SINR}{signal-to-interference-and-noise ratio}
    \acro{SNR}{signal-to-noise ratio}
    \acro{TOI}{target of interest}
    \acro{UE}{user equipment}
\end{acronym}

\bstctlcite{IEEEexample:BSTcontrol}

\title{Constellation Shaping for OFDM-ISAC Systems: From \hspace{-0.2em}Theoretical \hspace{-0.2em}Bounds \hspace{-0.2em}to \hspace{-0.2em}Practical \hspace{-0.2em}Implementation}

\author{Benedikt Geiger,~\IEEEmembership{Graduate Student Member, IEEE}, Fan Liu,~\IEEEmembership{Senior Member, IEEE}, Shihang Lu,~\IEEEmembership{Graduate Student Member, IEEE}, Andrej Rode,~\IEEEmembership{Graduate Student Member, IEEE}, Daniel Gil Gaviria,~\IEEEmembership{Graduate Student Member, IEEE}, Charlotte Muth,~\IEEEmembership{Graduate Student Member, IEEE}, Laurent Schmalen,~\IEEEmembership{Fellow, IEEE}        %
\thanks{This work has received funding from the German Federal Ministry of Research, Technology and Space (BMFTR) within the projects Open6GHub, Open6GHub+, KOMSENS-6G (grant agreements 16KISK010, 16KIS2405, and 16KISK123). Mr. Geiger acknowledges the Networking Grant from the Karlsruhe House of Young Scientists (KHYS). Parts of this paper have been presented at the IEEE International Symposium on Joint Communications and Sensing (JC\&S), Oulu, Finland, January 2025, in~\cite{Geiger25JCS}.}%
\thanks{Benedikt Geiger, Andrej Rode, Daniel Gil Gaviria, Charlotte Muth, and Laurent Schmalen are with the Communications Engineering Lab (CEL), Karlsruhe Institute of Technology (KIT), Hertzstr. 16, 76187 Karlsruhe, Germany (e-mail: first.lastname@kit.edu). Fan Liu is with the National Mobile Communications Research Laboratory, Southeast University, Nanjing 210096, China (e-mail: fan.liu@seu.edu.cn). Shihang Lu is with the School of Automation and Intelligent Manufacturing, Southern University of Science and Technology, Shenzhen 518055, China (e-mail: lush2021@mail.sustech.edu.cn).}
}

\markboth{DRAFT, February 2026}
{Shell \MakeLowercase{\textit{et al.}}: A Sample Article Using IEEEtran.cls for IEEE Journals}

\maketitle

\begin{abstract}
\Ac{ISAC} promises new use cases for mobile communication systems by reusing the communication signal for radar-like sensing. However, \ac{SC} impose conflicting requirements on the modulation format, resulting in a trade-off between their corresponding performance. This paper investigates constellation shaping as a means to simultaneously improve \ac{SC} performance in \ac{OFDM}-based \ac{ISAC} systems. We begin by deriving how the transmit symbols affect detection performance and derive theoretical lower and upper bounds on the maximum achievable information rate under a given sensing constraint. Using an autoencoder-based optimization, we investigate geometric, probabilistic, and joint constellation shaping, where joint shaping combines both approaches, employing both optimal maximum a-posteriori decoding and practical bit-metric decoding. Our results show that constellation shaping enables a flexible trade-off between \ac{SC}, can approach the derived upper bound, and significantly outperforms conventional modulation formats. Motivated by its practical implementation feasibility, we review \ac{PAS} and propose a generalization tailored to \ac{ISAC}. For this generalization, we propose a low-complexity log-likelihood ratio computation with negligible rate loss. We demonstrate that combining conventional and generalized \ac{PAS} enables a flexible and low-complexity trade-off between \ac{SC}, closely approaching the performance of joint constellation shaping.
\end{abstract}


\acresetall

\acresetall
\vspace{-0.2cm}
\section{Introduction}
\vspace{-0.15cm}

\IEEEPARstart{A}{utonomous} drones, intelligent transportation, and smart city infrastructures are novel applications that require precise environmental awareness seamlessly integrated with communications. Consequently, \ac{ISAC} has emerged as a key research direction for 6G, reusing communication signals for radar-like sensing of passive objects. \Ac{ISAC} is expected not only to reduce hardware redundancy compared to two separate systems but also to improve energy efficiency and overall system reliability of existing cellular networks~\cite{wild_6g_2023,liu_integrated_2022}.

Current 5G systems use \ac{OFDM} due to its numerous and significant advantages, e.g., high spectral efficiency, low-complexity equalization, easy rate adaptation, and easy multi-user downlink handling~\cite{OFDM_survey}. Beyond its role in communications, \ac{OFDM} offers also key advantages for radar systems. In particular, it has the lowest average ranging sidelobe for \ac{QAM} constellations~\cite{liu_cp-ofdm_2025} among all orthogonal signaling bases, and enables low-complexity range and velocity estimation of targets through \ac{FFT}-based processing~\cite{braun_ofdm_2014}. This makes \ac{OFDM} also a compelling choice for future 6G \ac{ISAC} systems~\cite{wild_6g_2023}. Fig.~\ref{fig:intro:ISAC_scenario} shows a sketch of the \ac{ISAC} scenario considered in this paper, which follows a standard setup from the literature~\cite{wild_6g_2023,keskin_fundamental_2025,du_probabilistic_2023}. The key idea is to use the same transmit signal of, e.g., the base station, for both communications to the \ac{UE} and sensing. In particular, the \ac{ISAC} transmit signal is used to convey information to a communication user, and to detect a potential target in a radar-like manner.

\begin{figure}[!t]
    \centering
    \input{figures/tikz_sensing_scenario}
    \vspace{-0.72cm}
    \caption{Considered \ac{ISAC} scenario. A base station employs a unified transmit signal with an optimized constellation to communicate with an \ac{UE} and to sense the environment in a radar-like manner. The objective of the \ac{ISAC} base station is to transmit data to the \ac{UE} and to detect a \ac{TOI}, such as a drone, in the presence of an interfering object like a building. Constellation shaping enables a flexible trade-off between communication throughput and detection probability of potential targets, and it improves both simultaneously compared with legacy modulation formats.}
    \label{fig:intro:ISAC_scenario}
    \vspace{-0.8cm}
\end{figure}

However, a fundamental difference between \ac{OFDM}-based radar and communication systems lies in the modulation format. \Ac{OFDM}-radar systems commonly use \ac{QPSK}~\cite{braun_ofdm_2014}, while communication systems employ higher-order modulation formats, such as \mbox{\num{64}-\ac{QAM}}, to increase spectral efficiency~\cite{keskin_fundamental_2025}. This is a result of the underlying trade-off between \ac{SC}, known as the \ac{DRT}~\cite{xiong_torch_2024,Shihang_dedicated_precoding}, which becomes particularly evident when considering the optimal channel input distributions for each task. To elaborate, a Gaussian distributed input maximizes the \ac{MI} for an \ac{AWGN} channel under an average power constraint, whereas constant modulus constellations achieve optimal sensing performance~\cite{xiong_torch_2024}. Consequently, both \ac{SC} performance are heavily influenced by the constellation. Therefore, constellation shaping was recently investigated for \ac{ISAC} to trade off \ac{MI} against sidelobe levels in the ambiguity function, aiming to balance \ac{SC} performance~\cite{du_probabilistic_2023, du_reshaping_2023}. While constellation design plays a critical role in communication performance, its influence on sensing-oriented metrics, e.g., detection probability, remains insufficiently explored. In practical \ac{OFDM}-\ac{ISAC} systems, it is still unclear how different constellations impact detection performance, or what trade-offs arise when balancing sensing accuracy with communication efficiency.

Constellation shaping, long studied in communications theory, can offer up to \SI{1.53}{dB} \ac{SNR} gain over uniform signaling~\cite{Forney_shaping_gain} in an \ac{AWGN} channel with an average power constraint. Yet, in the context of \ac{ISAC}, the fundamental limits remain unknown. In particular, the maximum achievable \ac{MI} constrained by sensing performance requirement has yet to be characterized.

Generally, there are three widely used approaches to constellation shaping:
\textit{Geometric shaping} optimizes the location of constellation points, assuming equal probability for each point~\cite{Gumus:20}. \textit{Probabilistic shaping} uses a conventional \ac{QAM} modulation format but optimizes the probability distribution of these points~\cite{Cho:19}. \textit{Joint shaping} simultaneously optimizes both the location and probability of the constellation points~\cite{stark_joint_2019}.

These different shaping methods have been successfully optimized and compared for communications using an \ac{AE} framework~\cite{stark_joint_2019,aref_end--end_2022}. Here, the communication system is modeled by differentiable blocks, and the constellation points and their probabilities are treated as trainable parameters. In contrast, within the context of \ac{ISAC}, only probabilistic constellation shaping has been studied in detail~\cite{du_probabilistic_2023, du_reshaping_2023,liu_probabilistic_2025}, and a systematic comparison between different shaping approaches is still missing. In particular, it remains an open question which shaping method is best suited for \ac{ISAC} systems, and how closely it can approach the maximum \ac{MI} under a given sensing constraint?

While previous work has addressed low-complexity constellation optimization~\cite{du_reshaping_2023}, the optimization itself can be done offline, with the resulting constellations stored as \acp{LUT}. Therefore, the key challenge is to design constellations that enable low-complexity implementation in practical systems.

Various schemes like Gallager's scheme, trellis shaping, or shell mapping have been proposed to bring probabilistic constellation shaping to practice, see e.g.,~\cite{bocherer_bandwidth_2015} and references therein. However, these often suffer from severe error propagation unless mitigated by computationally intensive algorithms. A key breakthrough was \ac{PAS}, which efficiently integrates \ac{FEC} into shaping, mitigating error propagation~\cite{bocherer_bandwidth_2015, buchali_rate_2016}. \ac{PAS} has become the state-of-the-art in commercial fiber-optic communication systems~\cite{sun_800g_2020} and is now under investigation for wireless applications~\cite{Gueltekin_ESS_wireless,PCS_5G}. However, conventional \ac{PAS} has been designed for communication systems, without considering sensing requirements. As a result, its structural constraints may limit the sensing performance, raising critical questions for \ac{ISAC}: How do the structural constraints of \ac{PAS} affect the trade-off between \ac{SC}? Can \ac{PAS} be generalized to improve \ac{SC} performance while preserving its low-complexity implementation advantage? How does the \ac{AIR} in practical \ac{ISAC} systems with \ac{BMD} compare to the \ac{MI}?\footnote{We note that the authors in~\cite{liu_probabilistic_2025} mention that constellation shaping can be implemented using \ac{PAS}, however they do not incorporate the structural constraints associated with \ac{PAS} into their optimization problem; the analysis is based on a loose upper bound on the \ac{MI} in the relevant operating regime; and performance is evaluated solely in terms of \ac{MI} overestimating the \ac{AIR} in practical systems with \ac{BMD}.}

In this paper, we extend our previous work~\cite{Geiger25JCS} investigating constellation shaping as a tool to improve \ac{SC} performance in a monostatic \ac{ISAC} system with a unified waveform for both \ac{SC}. Specifically, we optimize and systematically compare constellation shaping methods to maximize the \ac{AIR} under constraints on the sensing detection probability and the false alarm rate. The overarching goal of this work is, first, to establish fundamental performance limits and, second, to bridge the gap between theory and practice by demonstrating how these limits can be approached using low-complexity, practical systems. The key contributions of this paper are:
\vspace{-0.15cm}
\begin{itemize}
    \item \textbf{Theoretical characterization of the \ac{DRT}:} 
    We derive the constellation-dependent detection probability and show that it depends only on the kurtosis of the constellation. Based on this result, we derive lower and upper bounds on the maximum \ac{MI} given a detection probability constraint. This establishes a fundamental limit on the \ac{ISAC} performance and provides insight into the maximum achievable gain through constellation shaping.
    \item  \textbf{Autoencoder-based constellation optimization and systematic analysis:}
    We use a binary \ac{AE} to optimize constellations shaped geometrically, probabilistically, and jointly for both ideal receivers and practical systems with \ac{BMD}, while satisfying detection and false alarm constraints. We show that constellations optimized for ideal receivers closely approach the theoretical upper bound and quantify the performance loss when transitioning to practical \ac{BMD}. Furthermore, we demonstrate that joint constellation shaping outperforms geometric and probabilistic constellation shaping, enables flexible control over the \ac{SC} trade-off, and significantly outperforms conventional formats, e.g., 64-\ac{QAM}.
    \item \textbf{Low-complexity implementation of constellation shaping for practical \ac{ISAC} systems:} 
    We review \ac{PAS} and demonstrate its limitations in the \ac{ISAC} context. Therefore, we generalize \ac{PAS} to improve sensing performance and propose low-complexity \ac{LLR} computation with negligible \ac{AIR} loss. Simulation results show that a combination of conventional and generalized \ac{PAS} achieves \ac{SC} performance close to that of autoencoder-based optimization across a wide range of kurtosis constraints while remaining compatible with practical implementation requirements.
\end{itemize}

\textit{Notations:} Bold lowercase letters (e.g., $\vect{x}$) denote vectors, bold uppercase letters represent matrices (e.g., $\vect{J}$), and scalars are written in normal font (e.g., $N$). We do not explicitly distinguish between time and frequency domains; instead, the index implicitly indicates the respective domain: $t$ denotes continuous time, $k$ denotes discrete time/delay, and $n$ refers to the subcarrier index in the frequency domain. Sans-serif symbols (e.g.,~$\RV{x}$) indicate random variables, and calligraphic letters (e.g.,~$\mathcal{X}$) represent sets. The operator $(\cdot)^\star$ denotes complex conjugation. The Dirac delta function is written as $\delta(\cdot)$. The expectation with respect to the random variable $\RV{x}$ is denoted by $\expecv{\RV{x}}{\cdot}$. The functions $\mathbbm{h}(\cdot)$ and $\mathbb{H}(\cdot)$ denote the differential and discrete entropy, respectively. The real and imaginary parts of a complex number are denoted by $\Re\{\cdot\}$ and $\Im\{\cdot\}$, respectively. The operator $\mathrm{vec}(\cdot)$ denotes row-wise vectorization, and $\otimes$ denotes the outer product. The softmax function is written as $\mathrm{Softmax}(\cdot)$.

\begin{figure*}[!t]
    \centering
    \input{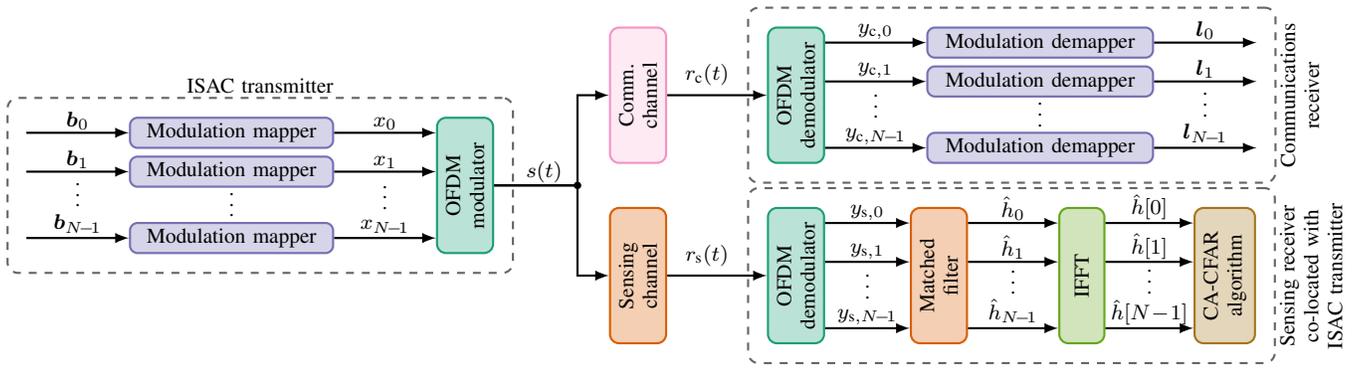}
    \vspace{-0.30cm}
    \caption{Block diagram of the considered monostatic \ac{OFDM}-\ac{ISAC} system. The unified \ac{ISAC} signal is employed for achieving both \ac{SC} tasks.}%
    \label{fig:system_model:block_diagram}
    \vspace{-0.5cm}
\end{figure*}

\vspace{-0.3cm}
\section{System model} \label{sec:ch2:system_model}
\vspace{-0.1cm}
In this work, we consider a monostatic \ac{OFDM}-\ac{ISAC} system as illustrated in Fig.~\ref{fig:system_model:block_diagram}. To reduce computational complexity and simplify the analysis, all $N$ sub-carriers use the same constellation, and the sensing target delays are modeled as integer multiples of the sampling period and are shorter than the duration of the \ac{CP}. Furthermore, we assume that the targets are static. We focus on the transmission of single \ac{OFDM} symbols, and neglect Doppler frequency, i.e., velocity, estimation.%

\vspace{-0.35cm}
\subsection{\acs{ISAC} Transmitter}
\vspace{-0.1cm}
For each sub-carrier \mbox{$n \in \{0, 1, \ldots, N \! - \! 1\}$}, a constellation mapper encodes $M$ bits \mbox{$\vect{b}_n \in \{0,1\}^{M}$}, onto a modulation symbol $x_n$ taken from the modulation alphabet $\mathcal{X}$, i.e., \mbox{$x_n \in \mathcal{X} \subset \mathbb{C}$}, where $\left| \mathcal{X} \right| = \widetilde{M} = 2^M$. Since the constellation symbols appear randomly selected and independent of each other, the transmit symbols $\RV{x}_n$ can be considered as i.i.d. \acp{RV} \mbox{$\RV{x}_n \sim P_{\RV{x}}(x_n)$}.

The \ac{OFDM} modulator transforms the $N$ frequency domain symbols $\RV{x}_n$, into the time domain using the orthonormal \ac{IFFT}. Next, a \ac{CP} is added before the signal $\RV{s}(t)$ is transmitted over an equivalent baseband channel.

\vspace{-0.35cm}
\subsection{Communications Channel and Receiver}
\vspace{-0.1cm}
For communications, the influence of a potential multi-path channel is assumed to be mitigated through equalization. 
The communication receiver removes the \ac{CP} from the received communication signal $\RV{r}_{\text{c}}(t)$ and converts it into the frequency domain using an orthonormal \ac{FFT}. This simplifies the \ac{OFDM} system into a set of $N$ parallel \ac{AWGN} channels, one for each sub-carrier. Therefore, the received communications symbol of the $n$th sub-carrier is given by
\vspace{-0.2cm}
\begin{equation}
    \vspace{-0.05cm}
    \RV{y}_{\text{c},n} = \RV{x}_n + \RV{w}_{\text{c},n},
    \label{eq:system_model:comm_channel}
    \vspace*{-0.25cm}
\end{equation}
where $\RV{w}_{\text{c},n} \sim \mathcal{CN}(0, \sigma_{\text{c,}n}^2)$ is \ac{AWGN} with variance $\sigma_{\text{c,}n}^2$.

At the receiver, we consider an optimal maximum a-posteriori decoder using \ac{SMD}, as well as \ac{BMD}, which is typically used in practical systems with binary error correcting codes~\cite{bocherer_bandwidth_2015}. In \ac{BMD}, a demapper computes a vector of \acp{LLR} $\vect{l}_n$, where the individual \acp{LLR} are evaluated separately for each bit position \mbox{$m = \{1, \ldots, M\}$}~\cite{Ivanov_BICM}
\vspace{-0.2cm}
\begin{equation}
    \vspace{-0.05cm}
    l_{n,m}(y_{\text{c},n}) \! = \!  \log \frac{\sum_{x_{\!n} \in \mathcal{X}^{(0)}_m}{ \! f_{\RV{y}_{\text{c,}n}|\RV{x}}(y_{\text{c},n}|x_n)}P_{\RV{x}}(x_n)}{\sum_{x_n \in \mathcal{X}^{(1)}_m}{ \! f_{\RV{y}_{\text{c,}n}|\RV{x}}(y_{\text{c},n}|x_n)}P_{\RV{x}}(x_n)},
    \label{eq:system_model:LLR}
    \vspace*{-0.25cm}
\end{equation}
where $\mathcal{X}^{(b)}_m$ represents the set of all constellation symbols labeled with bit \mbox{$b \in \{0,1\}$} at bit position $m$, and $f_{\RV{y}_{\text{c,}n}|\RV{x}}$ denotes the communications channel transition \ac{PDF}.%

\vspace{-0.2cm}
\subsection{Achievable Information Rates}
\vspace{-0.1cm}
The \ac{AIR} depends on the decoding metric, and we refer the reader to~\cite{bocherer_bandwidth_2015,buchali_rate_2016,fehenberger_probabilistic_2016} and the references therein for a comprehensive overview.

Under \ac{SMD}, where decoding is performed directly on the constellation symbols, the \ac{AIR} is given by the \ac{MI}~\cite{Cover2006}
\vspace{-0.2cm}
\begin{equation}
    \mathbb I_{\text{MI}}(\RV{x}_n; \RV{y}_{\text{c},n}) = \expecv{\RV{x}_n,\RV{y}_{\text{c},n}}{\log_2\left( \frac{f_{\RV{y}_{\text{c,}n}|\RV{x}}(\RV{y}_{\text{c},n}|\RV{x}_n)}{f_{\RV{y}_{\text{c,}n}}(\RV{y}_{\text{c},n})}\right)}, %
    \vspace*{-0.25cm}
\end{equation}
which defines the highest achievable rate with an arbitrarily complex receiver, where $f_{\RV{y}_{\text{c,}n}}(\RV{y}_{\text{c},n})$ denotes the \ac{PDF} of the received communication symbols $\RV{y}_{\text{c,}n}$ on sub-carrier $n$.

In contrast, practical communication systems employ \ac{BMD}, where the \ac{AIR} is lower bounded by the \ac{GMI}~\cite{bocherer_bandwidth_2015,fehenberger_probabilistic_2016}
\vspace{-0.2cm}
\begin{equation}
    \mathbb I_{\text{GMI}}(\bm{\RV{b}}_n; \RV{y}_{\text{c},n}) = \left[ \mathbb{H}(\bm{\RV{b}}_n) - \sum\limits_{m=1}^{M} \mathbb{H}(\RV{b}_{n,m}|\RV{y}_{\text{c},n}) \right]^+, %
    \label{eq:system_model:GMI}
    \vspace*{-0.25cm}
\end{equation}
where $[\; \cdot\;  ]^+$ is $\mathrm{max}(\cdot,\,0)$. The \ac{GMI} is upper bounded by the \ac{MI} $\mathbb I(\RV{x}_n; \RV{y}_{\text{c},n})$~\cite{Ivanov_BICM,fehenberger_probabilistic_2016}.

Consequently, we consider the \ac{MI} to explore the fundamental \ac{DRT} between \ac{SC} in idealized systems using \ac{SMD}, and use the \ac{GMI} to assess the \ac{SC} trade-off in practical communication systems employing \ac{BMD}.

\subsection{Sensing Channel and Receiver}
We consider a sensing scenario with $J$ targets, where one target is designated as the \ac{TOI}, e.g., a drone, and the remaining $J-1$ targets act as strong interferers, e.g., static buildings or residual self-interference from the finite isolation between the transmitter and receiver paths, which can be modeled as a static target at zero delay with an artificial \ac{RCS}. The \ac{RCS} of the \ac{TOI} follows a Swerling-1 model with \ac{PDF}
\vspace{-0.35cm}
\begin{equation}
    f_{\RV{\sigma}_{\text{RCS,TOI}}}(\sigma_{\text{RCS,TOI}})\! =\! \begin{cases} \!
        \frac{1}{\bar{\sigma}_{\text{RCS,TOI}}}
\exp\!\left(\!\!-\frac{\sigma_{\text{RCS,TOI}}}{\bar{\sigma}_{\text{RCS,TOI}}} \!\right)\!\!, \!\!\! \!& 
\sigma_{\text{RCS,TOI}} \! \geqslant \!0, \\
        \!0, & \sigma_{\text{RCS,TOI}}\! <\! 0,
    \end{cases}
    \vspace*{-0.18cm}
\end{equation}
i.e., the \ac{RCS} fluctuates independently from \ac{OFDM} symbol to \ac{OFDM} symbol according to an exponential distribution with mean $\bar{\sigma}_{\text{RCS,TOI}}$. The interfering targets follow a non-fluctuating Swerling-0 model and therefore have constant \ac{RCS} values $\sigma_{\mathrm{RCS},j}$, which are larger than the mean \ac{RCS} of the \ac{TOI}, i.e., \mbox{$\sigma_{\mathrm{RCS},j} \gg \bar{\sigma}_{\text{RCS}}$} for all \mbox{$j \neq \mathrm{TOI}$}. This detection setting is widely regarded as particularly challenging, since the sidelobes of the strong interfering targets can obscure the weak \ac{TOI}.
For each OFDM symbol, we draw one realization of the \ac{RCS} of the \ac{TOI} and treat it as fixed. Consequently, within a single symbol, the sensing channel can be modeled as the sum of $J$ static point targets with arbitrary, but fixed amplitudes
\vspace{-0.2cm}
\begin{equation}
    \RV{r}_{\text{s}}(t) = h(t) * \RV{s}(t) + \RV{w}_{\text{s}}(t) = \sum_{j=1}^{J} a_j \RV{s}(t - \tau_j) + \RV{w}_{\text{s}}(t), 
    \vspace{-0.2cm}
    \label{eq:system_model:sensing_channel_time_domain}
    \vspace*{-0.1cm}
\end{equation}
where $\RV{w}_{\mathrm{s}}(t) \sim \mathcal{CN}(0,\sigma_{\mathrm{s}}^2)$ denotes AWGN with variance $\sigma_{\text{s}}^2$. The amplitudes $a_j$ and delays $\tau_j$ follow from the radar equation
\vspace{-0.35cm}
\begin{equation}
        a_j = \sqrt{\frac{\sigma_{\mathrm{RCS},j} c_0^2 P_{\mathrm{Tx}} G_{\mathrm{Tx}} G_{\mathrm{Rx}}}
    {(4\pi)^3 R_j^4 f_{\mathrm{c}}^2}} \e^{\j \varphi_{j}},\qquad
    \tau_j = \frac{2 R_j}{c_0},
    \label{eq:system_model:target_amplitude_delay}
    \vspace*{-0.2cm}
\end{equation}
where $P_{\mathrm{Tx}}$, $G_{\mathrm{Tx}}$, $G_{\mathrm{Rx}}$, $f_{\mathrm{c}}$, and $c_0$ denote transmit power, antenna gains, carrier frequency, and speed of light. The phases $\varphi_j$ are independent across targets and, for each target, drawn i.i.d. from a uniform distribution over $[0,2\pi)$.
The sensing receiver samples the baseband signal $\RV{r}_{\text{s}}(t)$, removes the \ac{CP}, and transforms it into the frequency domain using the orthonormal \ac{FFT}. The frequency domain received symbols are
\vspace{-0.3cm}
\begin{equation}
    \RV{y}_{\text{s},n} = \RV{x}_n h_n + \RV{w}_{\text{s},n} = \RV{x}_n \sum_{j=1}^{J} a_j \e ^{- \j 2 \pi \frac{n}{N} \tau_j} + \RV{w}_{\text{s},n},
    \vspace{-0.18cm}
\end{equation}
where $h_n$ denotes the channel frequency response and $\RV{w}_{\text{s},n}$ is the \ac{FFT} of the sampled \ac{AWGN} $\RV{w}_{\text{s}}(t)$, and follows $\mathcal{CN} \left( 0, \sigma_{\text{s}}^2 \right)$.

The sensing receiver applies a sensing \ac{MF} 
\vspace{-0.1cm}
\begin{equation}
    \hat{\RV{h}}_n = \RV{y}_{\text{s},n} \RV{x}^{\star}_n = (h_n \RV{x}_n \! + \! \RV{w}_{\text{s},n} ) \RV{x}^{\star}_n = h_n \left| \RV{x}_n \right| ^2 \! + \! \RV{w}_{\text{s},n} \RV{x}^{\star}_n,  %
    \label{eq:system_model:transfer_function_estimate}
    \vspace{-0.1cm}
\end{equation}
which yields an unbiased estimate of the sensing channel
\vspace{-0.1cm}
\begin{equation}
    \mathbb{E}_{\RV{x},\RV{w}_{\text{s,}n}} \! \{\hat{\RV{h}}_n\} \! = \! \mathbb{E}_{\RV{x}} \{ h_n | \RV{x}_n  |^2  \} + \expecv{\RV{x},\RV{w}_{\text{s,}n}}{ \RV{w}_{\text{s},n} \RV{x}^{\star}_n } \! = \! h_n
    \label{eq:system_model:mean}
    \vspace{-0.1cm}
\end{equation} %
for unit power constellations $\mathbb{E}_{\RV{x}} \{ | \RV{x}_n  |^2  \} = \num{1}$.

The delay domain channel estimate $\hat{\RV{h}}[k]$ is obtained by applying the orthonormal \ac{IFFT} to the frequency domain channel estimate $\hat{\RV{h}}_n$
\vspace{-0.3cm}
\begin{equation}
    \hat{\RV{h}}[k]= \frac{1}{\sqrt{N}} \sum_{n=0}^{N-1} \hat{\RV{h}}_n \mathrm{e}^{\mathrm{j} 2 \pi \frac{n}{N} k}, \qquad k = \{0, \ldots N-1\}.
    \vspace*{-0.15cm}
\end{equation}

Finally, the \ac{CA}-\ac{CFAR} algorithm, which maximizes the detection probability $\PD$ given a maximum false alarm rate $\PFA$, determines whether a target is present at a delay $k$~\cite{richards_fundamentals_2014}.

\subsection{The Constellation-dependent Detection Probability} \label{sec:system_model:detection_probability}
We derive the constellation-dependent detection probability of the \ac{TOI} by analyzing each signal processing block of the sensing receiver. For the \ac{CA}-\ac{CFAR} algorithm with a window length \mbox{$N_{\text{win}} \rightarrow \infty$}, the detection probability is
\vspace{-0.1cm}
\begin{equation}
    \PD =  \PFA ^{\frac{1}{1 + \bar{\gamma}}_{\text{TOI}}},
    \label{eq:system_model:detection_prob}
    \vspace{-0.1cm}
\end{equation}
assuming Gaussian distributed noise and interference~\cite{richards_fundamentals_2014}. 
The detection probability $\PD$ depends only on the false alarm rate $\PFA$ and the average \ac{SINR} $\bar{\gamma}_{\text{TOI}}$ at the input of the detector. To prevent clutter-induced bursts of false alarms and to enable robust and predictable detection performance, the false alarm rate $\PFA$ is usually fixed during system design~\cite{richards_fundamentals_2014}. Consequently, increasing the detection probability $\PD$ requires increasing the average \ac{SINR} $\bar{\gamma}_{\text{TOI}}$ at the input of the \ac{CA}-\ac{CFAR}, i.e., at the output of the \ac{IFFT}. Note that the average is taken with respect to the \ac{RCS} fluctuations of the \ac{TOI}, as modeled by the Swerling-1 model. Following standard practice~\cite{richards_fundamentals_2014}, we first derive the \ac{SINR} for a fixed sensing-channel realization~(\ref{eq:system_model:sensing_channel_time_domain}) and subsequently average over the sensing-channel fluctuations.

\begin{figure}[!t]
	\centering
	\input{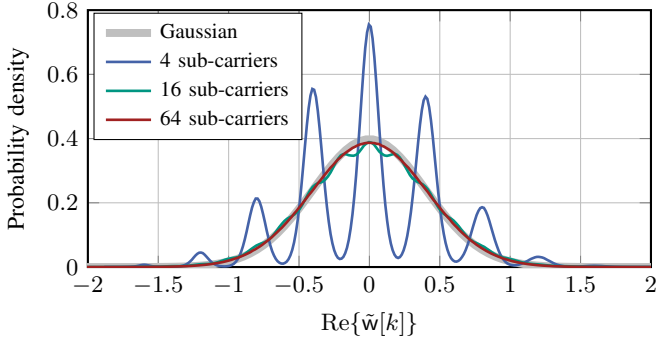}
    \vspace{-0.8cm}
	\caption{\Ac{PDF} of the noise at the input of the target detector, i.e., at the output of the \ac{IFFT} assuming a \num{16}-\ac{QAM} and a sensing \ac{SNR} of \SI{20}{dB} in a single target scenario for various numbers of sub-carriers.}
	\label{fig:system_model:noise_PDF}
    \vspace{-0.5cm}
\end{figure}

When random constellation symbols are transmitted, the noise at the output of the \ac{MF}, i.e., at the input of the \ac{IFFT}, may no longer be Gaussian as can be observed from~(\ref{eq:system_model:transfer_function_estimate}). To show that the noise is still approximately Gaussian at the output of the \ac{IFFT}, we decompose the channel transfer function estimate $\hat{\RV{h}}_n$ into a deterministic part $h_n$ and a random part \mbox{$\widetilde{\RV{w}}_n = h_n(|\RV{x}_n|^2 \!-\! 1)+\RV{w}_{\text{s,}n} \RV{x}^{\star}_n$}, which accounts for \ac{AWGN} and the randomness of the modulation, i.e., \mbox{$\hat{\RV{h}}_n = h_n + \widetilde{\RV{w}}_n$}.

For the random part $\widetilde{\RV{w}}_n$, the \ac{IFFT} acts as a summation of $N$ independent and scaled \acp{RV} $\widetilde{\RV{w}}_n$. According to the central limit theorem, the random part of the channel estimate \mbox{$\widetilde{\RV{w}}[k] = \hat{\RV{h}}[k] - h[k]$}, which is the \ac{IFFT} of $\widetilde{\RV{w}}_{n}$ approximates a Gaussian if the number of sub-carriers $N$ is sufficiently large. Fig.~\ref{fig:system_model:noise_PDF} shows the \ac{PDF} of $\Re \{\widetilde{\RV{w}}[k]\}$ for an increasing number of sub-carriers, demonstrating that the Gaussian assumption is already reasonable for \num{64} sub-carriers. This justifies modeling the detection probability using~(\ref{eq:system_model:detection_prob}).

Since the frequency-domain channel estimation errors are zero-mean and uncorrelated on different subcarriers, and using the linearity of the \ac{FFT}, the variance %
\vspace{-0.1cm}
\begin{equation}
\sigma^2_{\hat{\RV{h}}[k]} = \left( \frac{1}{\sqrt{N}} \right)^{\!2} \sum_{n = 0}^{N-1} \sigma^2_{\hat{\RV{h}}_n}
\label{eq:system_model:Variance_Delay_Frequency_domain}
\end{equation}
of the delay domain channel estimate equals the mean variance of the frequency domain channel estimates
\begin{align}
    \hspace*{-0.32cm} \sigma^2_{\hat{\RV{h}}_n} \! \! & = \mathbb{E}_{\RV{x},\RV{w}_{\text{s,}n}} \! \{ | \hat{\RV{h}}_n |^2  \} - | \mathbb{E}_{\RV{x},\RV{w}_{\text{s,}n}} \{\hat{\RV{h}}_n\} | ^2 \nonumber \\
    & = \mathbb{E}_{\RV{x},\RV{w}_{\text{s,}n}} \! \! \left\{ \! \left( \! h_n \!  \left| \RV{x}_n \right| ^2 \! \! \!+ \! \RV{w}_{\text{s},n} \RV{x}^{\star}_n \! \right) \! \! \left( \! h_n \! \left| \RV{x}_n \right| ^2 \!\! \!+\! \RV{w}_{\text{s},n} \RV{x}^{\star}_n \! \right)^{\! \! \star} \!\right\}\! \!  - \! \left| \! h_n \! \right|^{\! 2} \nonumber \\
    \begin{split}
        & = \mathbb{E}_{\RV{x},\RV{w}_{\text{s,}n}} \! \! \left\{ \left| h_n \right|^2 \left| \RV{x}_n \right |^4 \! + \! h_n \left| \RV{x}_n \right|^2 \RV{w}^{\star}_{\text{s,}n} \RV{x}_n \right. \\ & \qquad \qquad \; \left.+ \RV{w}_{\text{s,}n} \RV{x}^{\star}_n h_n^{\star} \left|\RV{x}_n \right|^2 \! +\!  \left| \RV{w}_{\text{s,}n} \RV{x}_n \right|^2  \right\} \! - \! \left| h_n \right|^2
    \end{split} \label{eq:system_model:variance_freq_domain} \\
 & = | h_n |^2 ( \mathbb{E}_{\RV{x}} \{ \left| \RV{x}_n \right|^4 \} - 1 )  +\sigma_{\mathrm{s}}^2  \nonumber \\
 & = | h_n |^2 \left( \kappa_n - 1 \right)  +\sigma_{\mathrm{s}}^2, \nonumber
\end{align}
where $\kappa_n$ is the kurtosis of the constellation
\vspace{-0.05cm}
\begin{equation}
    \kappa_n = \frac{\expecv{\RV{x}}{\left| \RV{x}_n - \expecv{\RV{x}}{\RV{x}_n}\right|^4}}{\left(\expecv{\RV{x}}{ \left| \RV{x}_n - \expecv{\RV{x}}{\RV{x}_n}\right|^2} \right)^2} \underset{\expecv{\RV{x}}{|\RV{x}_n|^2} = 1}{\overset{\expecv{\RV{x}}{\RV{x}_n} = 0}{=}} \expecv{\RV{x}}{|\RV{x}_n|^4},
    \vspace*{-0.05cm}
\end{equation}
which is equivalent to the 4th-order moment for unit power zero mean constellations. Moreover, since all subcarriers employ the same constellation, it follows that \mbox{$\kappa_n = \kappa$}. Inserting~(\ref{eq:system_model:variance_freq_domain}) into~(\ref{eq:system_model:Variance_Delay_Frequency_domain}) yields the noise-and-interference power
\vspace*{-0.2cm}
\begin{equation}
\sigma^2_{\hat{\RV{h}}[k]} =  \sum_{j=1}^{J}|a_j|^2 (\kappa-1) + \sigma_{\text{s}}^2,
\label{eq:system_model:noise_and_interference_power}
\vspace*{-0.2cm}
\end{equation}
where we exploit that all targets are independent with uniformly distributed phases
\vspace{-0.15cm}
\begin{equation}
    \frac{1}{N} \sum_{n=1}^{N-1}|h_n|^2 = \sum_{j=1}^{J} |a_j|^2.
    \vspace*{-0.3cm}
\end{equation}

For the deterministic part of the channel estimate, the \ac{IFFT} leads to an integration gain, increasing the power of the targets and consequently the \ac{SINR} by a factor of $N$, as
\vspace{-0.05cm}
\begin{equation}
    \frac{1}{\sqrt{N}} \sum_{n=0}^{N-1} \sum_{j=1}^{J} a_j \mathrm{e}^{-\mathrm{j} 2 \pi \frac{n \tau_j}{N}} \mathrm{e}^{\mathrm{j} 2 \pi \frac{n k}{N}} = \begin{cases}
        \sqrt{N} a_j, & k \! = \! \tau_j, \\
        0, & k \! \neq \! \tau_j.
    \end{cases}
    \label{eq:system_model:IFFT_deterministic}
    \vspace*{-0.05cm}
\end{equation}

The fraction of the power in the deterministic~(\ref{eq:system_model:IFFT_deterministic}) and random part~(\ref{eq:system_model:variance_freq_domain}) yields the \ac{SINR}
\vspace{-0.1cm}
\begin{equation}
    \gamma_{\mathrm{TOI}} = \frac{N \cdot | a_{\mathrm{TOI}} |^2}{ \sum_{j=1}^{J} |a_j |^2 ( \kappa -1 )  + \sigma_{\text{s}}^2 }
    \label{eq:system_model:SNR_TOI}
    \vspace*{-0.1cm}
\end{equation}
for a fixed sensing channel realization at the input of the \ac{CFAR} algorithm.

\begin{table}
\setlength{\tabcolsep}{4.5pt}
    \centering
    \caption{Kurtosis values of modulation formats}
    \vspace{-0.25cm}
    \label{tab:system_model:kurtosis}
    \begin{tabular}{c c c c c c}
        \toprule
        Constellation & $\widetilde{M}$-PSK & \num{16}-\ac{QAM} & \num{64}-\ac{QAM} & \num{256}-\ac{QAM} & $\mathcal{CN}(0, 1)$ \\
        \midrule
        Kurtosis $\kappa$ & \num{1} & \num{1.320} & \num{1.381} & \num{1.395} & \num{2} \\
        \bottomrule
    \end{tabular}
    \vspace{-0.4cm}
\end{table}
We observe that the \ac{SINR} depends only on the kurtosis $\kappa$
of the constellation for a given sensing channel realization. Using the identity \mbox{$\mathrm{Var}\{|\RV{x}|^2\} = \mathbb{E}\{|\RV{x}|^4\} - |\mathbb E\{|\RV{x}|^2\}|^2$} and the unit-power constraint \mbox{$\mathbb{E}\{|\RV{x}|^{2}\}=1$}, the kurtosis can be written as
\vspace{-0.25cm}
\begin{equation}
    \kappa = \mathrm{Var}\{|\RV{x}|^2\} + 1,
    \vspace*{-0.1cm}
\end{equation}
showing that the kurtosis directly measures the variance of the amplitudes of the constellation around the unit circle. Table~\ref{tab:system_model:kurtosis} lists the kurtosis of common modulation formats. Unit-modulus constellations such as \ac{PSK} achieve the minimum kurtosis $\kappa = 1$, preventing interference since the interference term in~(\ref{eq:system_model:noise_and_interference_power}) vanishes and only the \ac{AWGN} noise power $\sigma_{\mathrm{s}}^{2}$ remains. In contrast, conventional modulation formats such as $\widetilde{M}$-\ac{QAM} and a circular complex Gaussian \ac{PDF}, which maximizes the \ac{MI} for an \ac{AWGN} channel under an average transmit power constraint, introduce interference. For conventional $\widetilde{M}$-\ac{QAM}, the kurtosis approaches \mbox{$\kappa \approx 1.4$} as \mbox{$M \to \infty$} (see Appendix~\ref{sec:App_B}), while a circular complex Gaussian \ac{PDF} has \mbox{$\kappa = \num{2}$}.

We observe that the interference introduced by non-constant-modulus constellations, i.e., \mbox{$\kappa > 1$}, scales with the sum of the powers of all targets. Since the \ac{TOI} is assumed to be much weaker than the interfering targets, the interference power is dominated by the non-fluctuating interferers, such that the resulting noise-and-interference term is effectively independent of the \ac{RCS} fluctuations of the \ac{TOI}. In contrast, the received target power itself fluctuates according to the Swerling-1 model. Inserting~(\ref{eq:system_model:target_amplitude_delay}) into~(\ref{eq:system_model:SNR_TOI}) and averaging with respect to these fluctuations yields the average \ac{SINR} of the \ac{TOI}
\begin{equation}
    \bar{\gamma}_{\text{TOI}} =  \frac{N \cdot | \bar{a}_{\mathrm{TOI}} |^2}{ \sum_{j=1}^{J} |a_j |^2 ( \kappa -1 )  + \sigma_{\text{s}}^2 },
    \label{eq:system_model:mean_SNR_TOI}
\end{equation}
where $\bar{a}_{\mathrm{TOI}}$ denotes the average amplitude corresponding to the mean \ac{RCS} $\bar{\sigma}_{\text{RCS,TOI}}$. Substituting~(\ref{eq:system_model:mean_SNR_TOI}) into~(\ref{eq:system_model:detection_prob}) gives the constellation-dependent detection probability. Consequently, the \ac{RCS} fluctuations of the \ac{TOI} average out, and the detection probability depends only on the mean amplitude of the \ac{TOI}, the amplitude of the interfering targets, and the kurtosis $\kappa$ of the constellation.

\subsection{Optimization Problem}
In this paper, we aim to find constellations that maximize the communication performance, i.e., the (G)MI $\mathbb I_{\text{(G)MI}}$ of the overall \ac{ISAC} system, subject to a minimum detection probability constraint $\alpha_{\mathrm{D}}$, i.e., \mbox{$P_{\mathrm{D}} \geqslant \alpha_{\mathrm{D}}$}. 
Since all sub-carriers employ the same constellation, the total (G)MI is maximized if the (G)MI per sub-carrier is maximized. To ease optimization, we assume that all sub-carriers observe the same noise variance $\sigma^2_{\text{c}}$, i.e., \mbox{$\sigma^2_{\text{c,}n} = \sigma^2_{\text{c}}, \ \forall n$} and we omit the sub-carrier index $n$ in the following.\footnote{Frequency-selective fading results in a sub-carrier-dependent \ac{SNR}, which substantially complicates the optimization and obscures the fundamental \ac{SC} trade-off that we aim to characterize. If the transmitter has no \ac{CSI}, the shaped constellation would need to generalize over a broad range of \ac{SNR} values. Since the \ac{GMI}-optimal constellation depends on the operating \ac{SNR}, this would introduce not only the \ac{SC} trade-off studied in this work, but also an additional communications-only trade-off between low and high communication \ac{SNR} operation, thereby preventing a clean characterization of \ac{SC} trade-off. Conversely, if transmitter-side \ac{CSI} is available, adaptive modulation and power allocation, e.g., waterfilling, should be applied, which in turn affects the sensing performance and requires a joint optimization of constellation and power distribution. However, power allocation has only been explored recently~\cite{zhang_optimal_2025} and constitutes a research topic of its own. Including power allocation in this work would significantly complicate the optimization and obscure the shaping-induced trade-off, and is therefore beyond the focus of this paper. Consequently, we assume a constant noise power across the sub-carriers. This corresponds to an \ac{AWGN} communications channel and provides a clean first step for analyzing and characterizing the shaping-induced \ac{SC} trade-off, which is the focus of this work. Note that including the sub-carrier dependent \ac{SNR} into the optimization is an important direction for future work.}

Furthermore, we reformulate the detection-probability constraint \mbox{$P_{\mathrm{D}} \geqslant \alpha_{\mathrm{D}}$} as a kurtosis constraint \mbox{$\kappa \leqslant \tilde{\kappa}$}. This follows from the fact that the detection probability increases monotonically with the average \ac{SINR}~(\ref{eq:system_model:detection_prob}), and the average \ac{SINR} decreases monotonically with increasing kurtosis~(\ref{eq:system_model:mean_SNR_TOI}). Consequently, the detection probability $\PD$ is a monotonically decreasing function of the kurtosis $\kappa$ of the constellation, which directly justifies the reformulation of the detection-probability constraint as \mbox{$\kappa \leqslant \tilde{\kappa}$}. Although the average \ac{SINR}~(\ref{eq:system_model:mean_SNR_TOI}), and thus the detection probability, depend on the unknown target amplitudes $|a_j|^2$, these amplitudes are not required during the constellation optimization. We exploit the monotonic relationship between kurtosis and detection probability, which holds for any given realization of the target amplitudes. In other words, decreasing the kurtosis always improves the sensing performance for any possible set of target amplitudes, so that no prior knowledge of the unknown target amplitudes is required during the optimization.\footnote{The target amplitudes $|a_j|^2$ are required only to evaluate the resulting detection probability $P_{\mathrm{D}}$, but not to improve it by reducing the kurtosis $\kappa$.} This leads to the following optimization problem
\vspace{-0.15cm}
\begin{align}
    \max_{\mathcal{X}, P_{\RV{x}}(x)} \ & \mathbb I_{\text{(G)MI}} \label{eq:system_model:GMI_opt} \\
    \text{s.t. } & \textstyle \! \sum_{x}P_{\RV{x}}(x) = 1, \quad P_{\RV{x}}(x) \geqslant 0, \quad \forall x \in \mathcal{X}, \tag{C\num{0}} \label{eq:system_model:C0} \\
    & \mathbb{E}_{\RV{x}} \{ | \RV{x} |^2 \} = 1, \quad \; \; \mathbb{E}_{\RV{x}} \{ \RV{x} \} = 0, \tag{C\num{1}} \label{eq:system_model:C1} \\
    & \kappa \leqslant \tilde{\kappa}. \tag{C\num{2}} \label{eq:system_model:C2} %
    \vspace*{-0.45cm}
\end{align}

Here, the constraints~(\ref{eq:system_model:C0}) enforce that $P_{\RV{x}}(x)$ satisfies the properties of a probability mass function, while the constraints~(\ref{eq:system_model:C1}) ensure that the constellation has unit power \mbox{$E_{\text{s}} = \mathbb{E}_{\RV{x}}\{|\RV{x}|^2\} = 1$} and zero mean.

\vspace{-0.2cm}
\section{A Lower and Upper Bound for the maximum\\Mutual Information under a Kurtosis Constraint} \label{sec:bound}
In this section, we derive a lower and upper bound on the maximum \ac{MI} $\mathbb I(\RV{x}; \RV{y}_{\text{c}})$ under the constraints from~(\ref{eq:system_model:GMI_opt}). These bounds serve as performance benchmarks for our optimized constellations. To facilitate our study, we model the communication channel input as a continuous complex-valued \ac{RV} $\RV{x} \in \mathbb{C}$. A lower and upper bound on the \ac{MI}
\vspace{-0.15cm}
\begin{equation}
    \mathbb I(\RV{x};\RV{y}_{\text{c}}) = \mathbbm h(\RV{y}_{\text{c}}) - \mathbbm{h}(\RV{y}_{\text{c}}|\RV{x}) = \mathbbm{h}(\RV{y}_{\text{c}}) - \mathbbm{h}(\RV{w}_{\text{c}}) \label{eq:Bound:MI_Rewrite}
    \vspace*{-0.15cm}
\end{equation}
can be obtained by bounding the entropy of the received signal $\mathbbm{h}(\RV{y}_{\text{c}})$ as the entropy of the complex-valued \ac{AWGN} \mbox{$\mathbbm{h}(\RV{w}_{\text{c}}) =  \log_2 \!\left(\pi \mathrm{e} \sigma_{\text{c}}^2 \right)$} is constant~\cite{Cover2006}.

For the lower bound, we use the \ac{EPI}~\cite[Chap.~17.8]{Cover2006} expressed in bits
\vspace{-0.1cm}
\begin{equation}
    \mathbbm{h}(\RV{y}_{\text{c}}) \geqslant \log_2\left( 2^{\mathbbm{h}(\RV{x})} + 2^{\mathbbm{h}(\RV{w}_{\text{c}})}\right),
    \vspace*{-0.1cm}
\end{equation}
which is tight when $\RV{x}$ is Gaussian distributed.

Since we are interested in a lower and upper bound on the maximum \ac{MI}, we need to maximize the entropy of the transmit and received signal, respectively
\vspace{-0.1cm}
\begin{equation}
    \begin{split}
    \underbrace{\mathbbm{h}(\RV{y}_{\text{max,c}}) \! - \!  \mathbbm{h} (\RV{w}_{\text{c}})}_{\text{Upper bound}} \geqslant & \max_{f_{\RV{X}}(x): \text{\,(C0),\,(C1),\,(C2)}}  \mathbb I(\RV{x};\RV{y}_{\text{c}}) \\ \geqslant & \underbrace{\log_2\!  \left( 2^{\mathbbm{h}(\RV{x}_{\text{max}})} + 2^{\mathbbm{h}(\RV{w}_{\text{c}})}\right) \!- \! \mathbbm{h}(\RV{w}_{\text{c}})}_{\text{Lower Bound}} . \end{split}
    \label{eq:bound:final_bound}
\end{equation}
Here $\RV{x}_{\text{max}}$ and $\RV{y}_{\text{max,c}}$ denote the \acp{RV} associated with the entropy-maximizing \acp{PDF}.

In both cases, the entropy maximization problem has the same structure\footnote{Note that we omit the zero-mean constraint \mbox{$\mathbb{E}_{\RV{z}}\{z\} = 0$} here and show later that the entropy-maximizing \ac{PDF} satisfies it without explicit enforcement.}
\vspace{-0.2cm}
\begin{align}
    \max_{f_{\RV{z}}(z)} \ & \mathbbm{h} (\RV{z}) \\
    \text{s.t. } & \mathbb{E}_{\RV{z}} \{ | \RV{z} |^0 \} = C_0, \quad f_{\RV{z}}(z) \geqslant 0, \, \forall z \in \mathbb{C}, \tag{C\textsubscript{\num{0}}} \label{eq:bound:entropy_C0}  \\
    & \mathbb{E}_{\RV{z}} \{ | \RV{z} |^2 \} = C_1, \tag{C\textsubscript{\num{1}}} \label{eq:bound:entropy_C1} \\
    & \mathbb{E}_{\RV{z}} \{ | \RV{z} |^4 \} = C_2, \tag{C\textsubscript{\num{2}}} \label{eq:bound:entropy_C2}
    \vspace*{-0.2cm}
\end{align}
where $\RV{z}$ denotes either the transmit signal $\RV{x}$ (lower bound) or the received communication signal $\RV{y}_{\text{c}}$ (upper bound). The values of the corresponding moment constraints are summarized in Table~\ref{tab:bound:constraints}. For the upper bound, the transmit-side constraints on $\RV{x}$ are mapped into receive-side constraints on $\RV{y}_{\text{c}}$ and follow directly from substituting \mbox{$\RV{y}_{\text{c}} = \RV{x} + \RV{w}_{\text{c}}$} into the respective moment definitions.

\begin{table}
    \centering
    \caption{Moment constraints for entropy maximization}
    \vspace{-0.2cm}
    \label{tab:bound:constraints}
        \begin{tabular}{c c c c c}
        \toprule
        Bound & Domain $\RV{z}$ & $C_0$ & $C_1$ & $C_2$ \\
        \midrule
        Lower & Input $\RV{x}$ & $1$ & $E_{\text{s}}$ & $\tilde{\kappa}$ \\
        Upper & Output $\RV{y}_{\text{c}}$ & $1$ & $E_{\text{s}}+\sigma_{\text{c}}^2$ & $\tilde{\kappa} + 4E_{\text{s}}\sigma_{\text{c}}^2 + 2\sigma_{\text{c}}^4$ \\
        \bottomrule
    \end{tabular}
    \vspace{-0.5cm}
\end{table}

\textbf{Remark:} The optimization problem for the upper bound does not include any constraint ensuring that the resulting output \ac{PDF} is realizable by the system model. Specifically, there may not exist a transmit \ac{PDF} $f_{\RV{x}}$ such that passing it through the \ac{AWGN} channel yields the entropy-maximizing received \ac{PDF} $f_{\RV{y}_{\text{max,c}}}$. Consequently, the corresponding \ac{MI} represents a theoretical upper bound rather than an \ac{AIR}. Nevertheless, the upper bound gives insights into the gap between the achieved \ac{MI} of the optimized constellations and the theoretically maximum \ac{MI} under the given constraints.

According to the maximum entropy principle~\cite[Ch.~12.1]{Cover2006}, the \ac{PDF} that maximizes the entropy under moment constraints takes the exponential form. For constraints on the zeroth, second and fourth moments, the optimal \ac{PDF} has the form
\vspace{-0.25cm}
\begin{equation}
    f_{\RV{z}_{\text{max}}}(z) = \e^{\gamma_0 + \gamma_2 |z|^2 + \gamma_4 |z|^4},
    \label{eq:Bound:Ansatz_Maximum_entropy}
    \vspace*{-0.12cm}
\end{equation}
where the real-valued parameters $\gamma_0$, $\gamma_2$, and $\gamma_4$ are chosen to satisfy the constraint equations
\vspace{-0.15cm}
\begin{equation}
    C_q = \int_{\mathbb{C}} |z|^{2q} \e^{\gamma_0 + \gamma_2 |z|^2 + \gamma_4 |z|^4} \, \mathrm{d}z, \qquad q = 0,1,2.
    \label{eq:bound:non_linear_eq_system}   
    \vspace*{-0.1cm}
\end{equation}

We note that the optimal \ac{PDF} $f_{\RV{z}_{\text{max}}}(z)$ is circularly symmetric with respect to the origin and, as a consequence, has zero mean, thus fulfilling constraint~(\ref{eq:system_model:C1}). The resulting system of equations can be solved numerically. However, this system involves integrals over the complex plane and comprises three coupled equations in the three unknowns $\gamma_0$, $\gamma_2$, and $\gamma_4$, which leads to a comparatively high computational complexity. To reduce computational complexity, we reformulate the system of equations~(\ref{eq:bound:non_linear_eq_system}) in Appendix~\ref{sec:App_A} such that $\gamma_2$ can be obtained by numerically solving a single nonlinear equation
\vspace{-0.15cm}
\begin{align}
    \begin{split}
            C_0 =&   \left( \frac{C_1(\gamma_2C_1 \!+\!1)}{C_2}-\gamma_2\right)\sqrt{\frac{\pi C_2}{2(\gamma_2C_1\!+\!1)}} \\
            & \cdot \exp\! \left(\frac{\gamma^2_2C_2}{2(\gamma_2C_1\!+\!1)}\right) \mathrm{erfc}\!\left(\! \! -\gamma_2\sqrt{\frac{C_2}{2(\gamma_2C_1\!+\!1)}} \right).
            \label{eq:bound:gamma2_1D}
    \end{split}
\end{align}
Then, $\gamma_0$ and $\gamma_4$ follow immediately in closed form
\vspace{-0.22cm}
\begin{align}
    \gamma_0 & = \ln \! \left(\frac{1}{\pi} \left[ \frac{C_1(\gamma_2 C_1 +1)}{C_2} - \gamma_2\right] \right), \label{eq:bound:gamma0_1D} \\
    \gamma_4 & = -|\gamma_4| = \frac{-1}{2C_2} (\gamma_2 C_1 + 1). 
    \label{eq:bound:gamma4_1D}
\end{align}
Consequently, determining $\gamma_2$ is the key step and can be achieved by solving a single nonlinear equation~(\ref{eq:bound:gamma2_1D}) numerically using standard one-dimensional root-finding methods.

Note that for \mbox{$\kappa = 2$}, the optimal distribution reduces to a circular complex Gaussian, which corresponds to \mbox{$\gamma_2 = -1/C_1$} and \mbox{$\gamma_4 = 0$}. Starting from this point, we can continuously and smoothly transition from the Gaussian case (\mbox{$\kappa = 2$}) to a unit-modulus distribution (\mbox{$\kappa = 1$}) by gradually shifting probability mass toward the unit circle (see Fig.~\ref{fig:results:distribution_received_signal}), achieved by increasing \mbox{$\gamma_2$} and decreasing \mbox{$\gamma_4$}. This continuous transition ensures that every target value \mbox{$\kappa \in (1,2]$} is attainable along this path. Moreover, if there exists a solution for the entropy-maximizing distribution, it is unique~\cite{Cover2006}. Since~(\ref{eq:bound:gamma2_1D}) is a reformulation of the corresponding moment-matching condition~(\ref{eq:bound:non_linear_eq_system}), the solution for $\gamma_2$ is likewise unique for the considered kurtosis range \mbox{$\kappa \in [1,2]$}, apart from potential numerical instabilities caused by vanishing or exploding terms. In such cases, the system of equations~(\ref{eq:bound:non_linear_eq_system}) can be solved instead. This does not pose a practical issue, since the bounds are computed offline and are used solely as a reference for evaluating the optimized constellations.

\begin{figure*}[!t]
    \centering
    \input{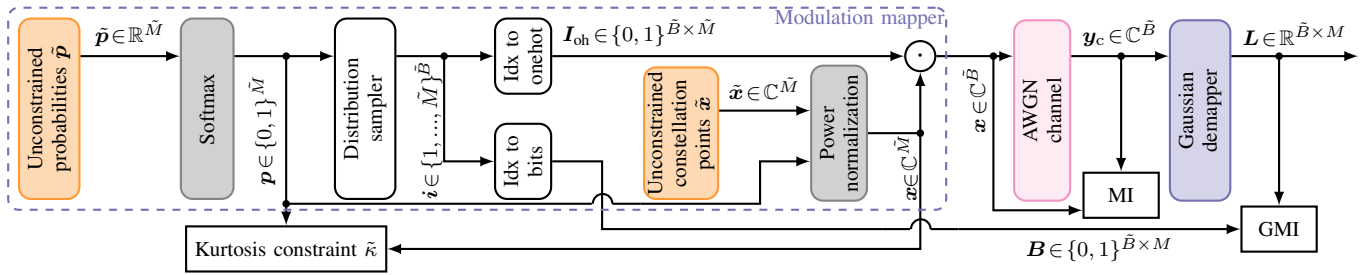}
    \vspace{-0.75cm}
    \caption{Proposed \ac{AE} framework to shape constellations for \ac{ISAC}. The trainable parameters are marked in orange and their normalization in gray.}
    \label{fig:AE:setup}
    \vspace{-0.6cm}
\end{figure*}

Once the parameters for the transmit and receive \acp{PDF} are known, the respective entropy $\mathbbm{h}(\RV{z}_{\text{max}})$ is 
\vspace{-0.25cm}
\begin{equation}
    \mathbbm{h}(\RV{z}_{\text{max}})  = \frac{-\gamma_0 - C_1 \cdot \gamma_2 - C_2 \cdot \gamma_4}{\ln(2)},
    \label{eq:bounds:entropy_maximizing_distribution}
    \vspace*{-0.2cm}
\end{equation}
which follows immediately by inserting~(\ref{eq:Bound:Ansatz_Maximum_entropy}) into the definition of the entropy.

To obtain the lower and upper bounds, the following steps must be carried out separately for each bound. First, compute the constraints $C_0$, $C_2$, and $C_4$ in Table~\ref{tab:bound:constraints}. Second, solve~(\ref{eq:bound:gamma2_1D}) numerically to obtain $\gamma_2$. Third, compute $\gamma_0$~(\ref{eq:bound:gamma0_1D}) and $\gamma_4$~(\ref{eq:bound:gamma4_1D}). Fourth, determine the entropy~(\ref{eq:bounds:entropy_maximizing_distribution}). Finally, insert this entropy into~(\ref{eq:bound:final_bound}) to obtain the respective bound. We provide source code to reproduce these bounds in~\cite{source_code_isac}.

\textbf{Remark:} Both bounds are tight for \mbox{$\kappa = 2$}, where they coincide with the Shannon capacity of an \ac{AWGN} channel \mbox{$\log_2(1 + E_{\text{s}}/\sigma^2_{\text{s}})$}. For \mbox{$\kappa < 2$}, the bounds are generally loose: the lower bound is tight only when $\RV{x}$ is Gaussian, and the upper bound may not be attainable because the entropy-maximizing receive \ac{PDF} $f_{\RV{y}_{\text{max,c}}}$ is not realizable as discussed in the previous remark. Nevertheless, the bounds become asymptotically tight as the communication \ac{SNR} increases, since the noise power $\sigma^2_{\text{c}}$ decreases and the constraints $C_1$ and $C_2$ in Tab.~\ref{tab:bound:constraints} converge. Moreover, our numerical results in Sec.~\ref{sec:simulation} and Fig.~\ref{fig:results:random_determinisitic} show that the \ac{MI} of the optimized constellations approaches the upper bound closely.

\vspace{-0.3cm}
\section{Constellation Shaping using Autoencoders} \label{sec:AE}
\vspace{-0.15cm}

In this section, we propose to incorporate the sensing constraint~(\ref{eq:system_model:C2}) into the bitwise \ac{AE} framework~\cite{stark_joint_2019}, enabling the joint end-to-end optimization of both geometric and probabilistic shaping for \ac{SC}. Autoencoders have become a standard and state-of-the-art tool for constellation shaping in communication systems, see, e.g., \cite{OShea_Deep_Learning_Physical_Layer, cammerer_trainable_2020}. Although other machine-learning-based optimization techniques could in principle be used, we are not aware of any approach that systematically outperforms \ac{AE}-based end-to-end optimization for constellation shaping. A further key advantage of the \ac{AE} approach is that it naturally supports probabilistic, geometric, and joint shaping within a unified architecture, whereas many existing methods focus on only one of these aspects.

A classical \ac{AE} consists of an encoder that learns an internal representation of the input and a decoder that attempts to reconstruct the input, typically both implemented as neural networks. When applied to communication systems, in par\-ti\-cu\-lar to constellation shaping, the encoder takes the role of the transmitter by learning a representation of the binary input in the form of an optimized constellation, while the decoder takes the role of the receiver by recovering the transmitted bits. During training, encoder and decoder are optimized jointly in an end-to-end manner such that the \ac{AIR} is maximized. For more details on \ac{AE}-based constellation optimization, we refer the reader to~\cite{aref_end--end_2022, stark_joint_2019, rode_end--end_2023}.

The block diagram of the \ac{AE} is shown in Fig.~\ref{fig:AE:setup}. In our setup, the receiver is a demapper with a Gaussian noise assumption~\cite{Ivanov_BICM}, and only the constellation of the transmitter is trainable. Depending on the shaping method (geometric, probabilistic, or joint), we optimize the constellation points, their probabilities, or both. For geometric shaping, only the unconstrained constellation points $\tilde{\vect{x}}$ are trainable. The unconstrained probabilities $\tilde{\vect{p}}$ are initialized uniformly and remain fixed, i.e., they are not trainable. Thus, the \ac{AE} optimizes the position of the constellation points while the probability of occurrence remains unchanged, i.e., uniform. For probabilistic shaping, the constellation points are initialized, e.g., as conventional $\widetilde{M}$-\ac{QAM}, and remain fixed during training, while only the unconstrained probabilities $\tilde{\vect{p}}$ are trainable. This corresponds to optimizing the probability of occurrence of the constellation points while keeping their position fixed. For joint, i.e., both geometric and probabilistic, constellation shaping, the unconstrained constellation points $\tilde{\vect{x}}$ and the unconstrained probabilities $\tilde{\vect{p}}$ are trainable, which yields the most degrees of freedom. The trainable parameters, shown as orange blocks in Fig.~\ref{fig:AE:setup}, are implemented as linear layers without bias.

\begin{figure*}[!t]
    \centering
    \input{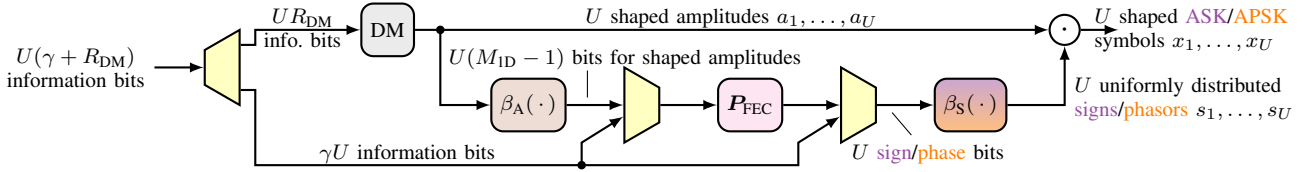}
    \vspace{-0.4cm}
    \caption{Encoding structure of conventional \ac{PAS}, based on~\cite{bocherer_bandwidth_2015, buchali_rate_2016}, and the proposed generalized \ac{PAS}. In both schemes, a distribution matcher (DM) generates shaped amplitudes, while sign or phase bits are derived from a combination of uniformly distributed parity bits $\vect{P}_{\text{FEC}}$ and information bits. In conventional \ac{PAS}, a single sign bit modulates the sign of real-valued \ac{ASK} symbols, requiring two parallel encoders for the real and imaginary parts. In generalized \ac{PAS}, multiple phase bits are mapped to discrete phases, directly generating complex-valued, circularly symmetric constellations. The generalization preserves the advantages of the original scheme while enhancing sensing performance.}
    \label{fig:PAS:encoding_scheme}
    \vspace{-0.65cm}
\end{figure*}

For a batch size $\tilde{B}$, the distribution sampler generates random indices \mbox{$\vect{i} \in \{1, \ldots, \tilde{M} \}^{\tilde{B}}$} according to the input distribution \mbox{$P_{\RV{x}}(x) \triangleq \vect{p}$}. To ease optimization, we use the Gumbel-softmax trick and optimize the unconstrained probabilities \mbox{$\tilde{\vect{p}} \in \mathbb{R}^{\tilde{M}}$} which are passed through a softmax layer to ensure that $\vect{p}$ satisfies the probability distribution properties~(\ref{eq:system_model:C0})~\cite{stark_joint_2019}.

Then, these indices are mapped to bits~$\vect{B}$ and one-hot vectors~$\vect{I}_{\mathrm{oh}}$. To select the transmit symbols, the one-hot vectors are multiplied with the normalized constellation points $\vect{x}$. Similar to the probabilities, we optimize the unconstrained constellation points $\tilde{\vect{x}}$, which are passed through a power normalization layer to obtain the unit power constellation points \mbox{$\mathcal{X} \triangleq \vect{x}$}.

Next, the selected constellation symbols are transmitted over an \ac{AWGN} channel~(\ref{eq:system_model:comm_channel}). For \ac{BMD}, the \acp{LLR}~(\ref{eq:system_model:LLR}) are computed for each bit using a classical Gaussian demapper~\cite{Ivanov_BICM}. The resulting \ac{GMI}~(\ref{eq:system_model:GMI}) is computed from the bits $\vect{B}$ and \acp{LLR} $\vect{L}$~\cite{fehenberger_probabilistic_2016}. For \ac{SMD}, the \ac{MI} is directly computed from the transmit symbols $\vect{x}$ and receive symbols $\vect{y}_{\text{c}}$.

We showed in Sec.~\ref{sec:ch2:system_model} that the detection probability constraint can be reformulated as a kurtosis constraint~(\ref{eq:system_model:C2}). Therefore, we propose a sensing loss term that depends on the kurtosis $\kappa$ of the constellation to satisfy the detection probability constraint
\vspace{-0.4cm}
\begin{equation}
    L_{\text{sens}} = {\begin{cases}
        0, &   \kappa \leqslant \tilde{\kappa}, \\
        d (\kappa - \tilde{\kappa}), & \kappa > \tilde{\kappa}.
    \end{cases}}
    \label{eq:AE:sensing_loss}
    \vspace{-0.15cm}
\end{equation}
The penalty factor \mbox{$d \in \mathbb{R}^{+}$} controls the strength of the penalty if the kurtosis threshold $\tilde{\kappa}$ is violated. To improve \ac{SC} performance simultaneously, the overall non-negative loss function combines both \ac{SC} performance
\vspace{-0.1cm}
\begin{equation}
    L_{\text{S\&C}} = \underbrace{\frac{M - \mathbb I_{\text{(G)MI}}}{M}}_{\text{Communications}} + L_{\text{sens}}.
    \label{eq:AE:overall_loss}
    \vspace{-0.1cm}
\end{equation}
Note that this loss function does not strictly enforce the sensing constraint. However, the communications loss term is normalized between $[0,1]$ and by selecting the penalty factor $d$ sufficiently large, the sensing loss term dominates if the kurtosis constraint is violated, effectively enforcing the sensing constraint.

\vspace{-0.2cm}
\section{Low-Complexity Implementation of Constellation Shaping for ISAC}
In the previous sections, we focused on the theoretical limits and the optimization procedure for constellation shaping without addressing its practical implementation. A promising candidate to implement probabilistic constellation shaping is \ac{PAS}, which integrates shaping and \ac{FEC} in a structured, hardware-efficient manner, effectively mitigating error propagation. We begin this section by revisiting \ac{PAS}, referring the reader to~\cite{bocherer_bandwidth_2015, buchali_rate_2016} for a detailed introduction. Our focus lies on how the \ac{PAS} structure constrains constellation design, and how these constraints can be incorporated into our \ac{AE} framework to optimize \ac{PAS}-compatible constellations. Then, we show how \ac{PAS} enables low-complexity \ac{LLR} computation at the receiver. %

\vspace{-0.3cm}
\subsection{Conventional Probabilistic Amplitude Shaping}
\vspace{-0.10cm}

\subsubsection{Constellation Design and Encoding Scheme} 
Conventional \ac{PAS} constructs a complex-valued constellation point by independently shaping and combining two real-valued $2^{M_{\text{1D}}}$-\ac{ASK} components, where \mbox{$M_{\text{1D}} = M/2$}. Fig.~\ref{fig:PAS:encoding_scheme} shows the encoding procedure for a single real-valued dimension corresponding to one of the two components of a complex-valued modulation mapper in Fig.~\ref{fig:system_model:block_diagram}.

\ac{PAS} generates blocks of \mbox{$U \in \mathbb{N}$} \ac{ASK} symbols with target amplitude distribution $P_{\RV{a}}(a_u)$. To this end, the information bits are split into two parts. The first fraction, consisting of \mbox{$U \cdot R_{\text{DM}}$} bits, is passed through a \ac{DM}, e.g., a constant composition \ac{DM}~\cite{bocherer_bandwidth_2015}. The rate $R_{\text{DM}}$ asymptotically approaches $\mathbb{H}(\RV{a})$ and we assume that \mbox{$U \cdot R_{\text{DM}}$} is integer~\cite{bocherer_bandwidth_2015}. The \ac{DM} maps the bits onto $U$ shaped amplitudes \mbox{$a_u \in \mathcal{A} = \Delta \cdot \{1, 3, \ldots, 2^{M_{\text{1D}} - 1} -1\}$}, for \mbox{$u \in \{0, 1, \ldots, U-1\}$}, following the target amplitude distribution $P_{\RV{a}}(a_u)$, where $\Delta$ is a normalization factor. Each amplitude $a_u$ is labeled using \mbox{$(M_{\text{1D}} - 1)$} bits via binary reflected Gray coding \mbox{$\beta_{\text{A}}: \mathcal{A} \rightarrow \{0,1\}^{M_{\text{1D}} - 1}$}.

The remaining $\gamma U$ information bits are first appended to the \mbox{$U(M_{\text{1D}}-1)$} bits representing the $U$ shaped amplitudes, where \mbox{$\gamma > 0$} controls how many bits bypass the \ac{DM}. The combined bitstream is then encoded using a systematic \ac{FEC} code with rate \mbox{$(M_{\text{1D}}-1+\gamma)/M_{\text{1D}}$}. Since the encoder is systematic, the shaped amplitude distribution is preserved at the output. The resulting parity bits, which are approximately uniformly distributed~\cite{bocherer_bandwidth_2015}, are next combined with the $\gamma U$ information bits to form the $U$ sign bits. These determine the signs \textcolor{Set1-D}{\mbox{$s_u \in \mathcal{S}_{\text{CPAS}} = \{-1, +1\}$}} of the \ac{ASK} symbols via the mapping \textcolor{Set1-D}{\mbox{$\beta_{\text{S}}: \{0,1\} \rightarrow \mathcal{S}_{\text{CPAS}}$}}.

Consequently, each one-dimensional \ac{ASK} transmit symbols is constructed via amplitude-sign factorization \mbox{$x_{\text{1D},u} = a_u \cdot s_u$} and is represented by one sign bit and \mbox{$(M_{\text{1D}} \!\!-\!\! 1)$} amplitude bits.

Since shaping is applied only to amplitudes and the sign bits remain uniform, this structure imposes a symmetry constraint on the one-dimensional \ac{ASK} distribution, i.e.,
\vspace{-0.2cm}
\begin{equation}
P_{\RV{x}_{\text{1D}}}(x_{\text{1D}}) = P_{\RV{x}_{\text{1D}}}(-x_{\text{1D}}), \quad \forall x_{\text{1D}} \in \mathcal{X}_{\text{1D}},
\end{equation}
where \mbox{$\mathcal{X}_{\text{1D}} = \{ s \cdot a : s \in \mathcal{S}_{\text{CPAS}},\, a \in \mathcal{A} \}$}. The same one-dimensional distribution is typically used for both dimensions of the complex constellation~\cite{bocherer_bandwidth_2015,buchali_rate_2016}.

\begin{figure*}[!t]
    \centering
    \hspace*{-0.6cm}
    \input{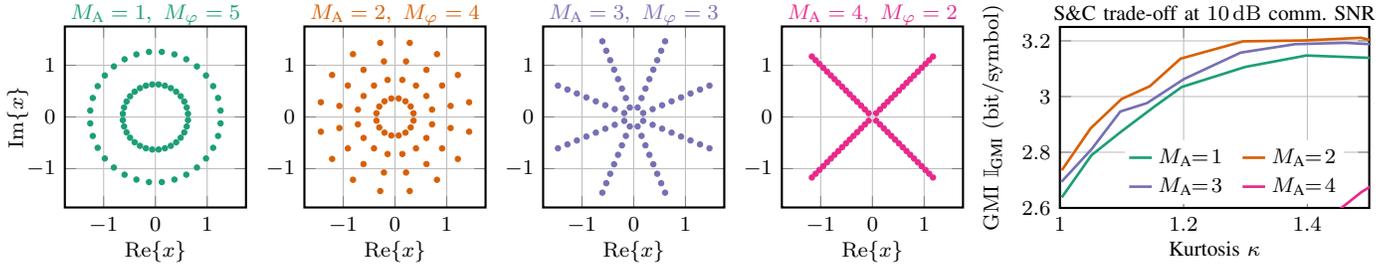}
    \vspace{-0.4cm}
    \caption{Generalized \ac{PAS} constellations for different combinations of $M_{\text{A}}$ and $M_\varphi$ with $M = 6$ total bits, and their corresponding \ac{SC} trade-off.}
    \label{fig:GPAS:Constellation_Examples}
    \vspace{-0.5cm}
\end{figure*}

\subsubsection{Optimizing Conventional PAS-compatible Constellations using an AE framework}
In general probabilistic shaping, the \ac{AE} optimizes $\widetilde{M}$ unconstrained probabilities $\tilde{\vect{p}}$ directly (see Sec.~\ref{sec:AE} and Fig.~\ref{fig:AE:setup}). However, as discussed in the previous paragraph, conventional \ac{PAS} imposes a specific structure on the probability of occurrence of the constellation points: (i) the probability of a complex-valued symbol factorizes into one-dimensional probabilities for the in-phase and quadrature components, and we enforce identical marginal distributions to reduce demapping complexity, (ii) the amplitude distribution is symmetric to enforce that the sign bits are uniformly distributed. Thus, to optimize \ac{PAS}-compatible constellations using the \ac{AE} framework, this structure must be enforced inside the \ac{AE}. Instead of learning all $\widetilde{M}$ unconstrained probabilities $\tilde{\vect{p}}$, only the unconstrained parameters $\tilde{\vect{p}}_{\RV{a}}$, which define the one-dimensional target amplitude distribution
\mbox{$\mathrm{Softmax} ( \tilde{\vect{p}}_{\RV{a}}) = [ P_{\RV{a}}(a_1), \ldots, P_{\RV{a}}(a_{2^{M_{\text{1D}}-1}})]$}, are learned. Based on this optimized target amplitude distribution, the probability of occurrence of all constellation points is then generated.
To impose symmetry, we construct an unconstrained, symmetric one-dimensional probability vector 
\vspace{-0.2cm}
\begin{equation}
    \tilde{\vect{p}}_{\RV{x}_{\text{1D}}} = [\tilde{\vect{p}}_{\RV{a}}, \vect{J}_{\text{N}} \tilde{\vect{p}}_{\RV{a}}] \, \in \, \mathbb R^{\left(2^{M_{\text{1D}}}\right)},
    \vspace{-0.15cm}
\end{equation}
by concatenating $\tilde{\vect{p}}_{\RV{a}}$ with its flipped version, where $\vect{J}_{\text{N}}$ denotes the anti-identity matrix that reverses the order of the vector entries. The final two-dimensional probability mass vector $\vect{p}_{\text{CPAS}}$ is obtained by applying softmax to each dimension and taking the outer product, followed by row-wise vectorization
\vspace{-0.2cm}
\begin{equation}
\vect{p}_{\text{CPAS}} = \mathrm{vec} \left(
\mathrm{Softmax}(\tilde{\vect{p}}_{\RV{x}_{\text{1D}}})
\otimes
\mathrm{Softmax}(\tilde{\vect{p}}_{\RV{x}_{\text{1D}}})
\right) \, .
\vspace{-0.15cm}
\end{equation}
The probability mass vector $\vect{p}_{\text{CPAS}}$ satisfies the constraints of a conventional \ac{PAS} constellation and replaces the probabilities $\vect{p}$ in the \ac{AE} shown in Fig.~\ref{fig:AE:setup}. This also significantly reduces the number of trainable parameters: instead of $\widetilde{M}$ unconstrained probabilities, the $2^{M_{\text{1D}}-1}$ unconstrained parameters of the one-dimensional target amplitude distribution $\tilde{\vect{p}}_{\RV{a}}$ define the probability of occurrence of all constellation points.

\subsubsection{Low-complexity LLR Calculation} \label{sec:pas:LLR}
With \ac{PAS}, the receiver still uses classical \ac{BMD}, which relies on~(\ref{eq:system_model:LLR}). However, while~(\ref{eq:system_model:LLR}) can be easily evaluated offline, i.e., during the optimization of the constellations, it is computationally too complex for a real-time implementation in 6G systems. Therefore, we briefly discuss how the structural properties of \ac{PAS} can be exploited to reduce the computational complexity. Since the constellation points and their probabilities are fixed after optimization, the \acp{LLR} can be precomputed and stored in \acp{LUT}. However, in the general case, this requires a separate two-dimensional \ac{LUT} per bit position, which covers both real and imaginary components, resulting in prohibitive memory requirements. Here, the structure of \ac{PAS} offers a key advantage: due to the independent modulation of the real and imaginary parts and the separability of the channel~\cite{fehenberger_probabilistic_2016}, the \acp{LLR} can be computed separately per dimension and stored in one-dimensional \acp{LUT}, yielding low-complexity, memory-efficient \ac{LLR} computation suitable for practical 6G receivers.

\vspace{-0.3cm}
\subsection{Generalized Probabilistic Amplitude Shaping}
In conventional \ac{PAS}, the real and imaginary components are shaped independently, and the joint distribution is the Cartesian product of two one-dimensional marginals. However, constructing \ac{PSK}-like constellations, which are preferred under strong sensing constraints, from such a product is not possible. As a result, the Cartesian structure of conventional \ac{PAS} inherently limits circular symmetry, resulting in sub-optimal \ac{SC} performance under strong kurtosis constraints. %

To overcome this limitation, we propose a generalization of the conventional \ac{PAS} scheme. In particular, we extend the original \ac{PAS} concept from \ac{ASK} to \ac{APSK}-like constellations, resulting in circularly symmetric constellations that are better suited for sensing. Furthermore, we show that the corresponding \acp{LLR} can still be approximated with low complexity and minimal memory, thereby preserving the implementation benefits of the original \ac{PAS} architecture.

\begin{figure*}[!t]
    \input{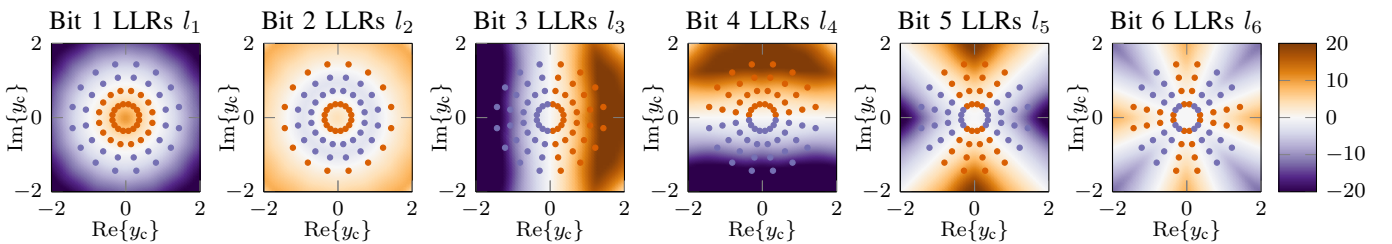}
    \vspace{-0.65cm}
    \caption{Two-dimensional \ac{LLR} values of received symbols for generalized \ac{PAS} with a total of $M = 6$ bits, composed of $M_{\text{A}} = 2$ amplitude bits (Bits 1 and 2) and $M_\varphi = 4$ phase bits (Bits 3 to 6), assuming equal probability of all constellation points at an \ac{SNR} of \SI{10}{dB}. The constellation points are colored orange if the corresponding bit is 0 and purple if it is 1.}
    \label{fig:PAS:GLLR_values_2D}
    \vspace{-0.05cm}
\end{figure*}

\subsubsection{Constellation Design and Encoding Scheme}
Conventional \ac{PAS} operates on \ac{ASK} constellations, where a single bit denotes the sign. We generalize this concept by introducing multiple “sign” bits, interpreted as phase bits, which no longer yield bipolar constellations, as in \ac{ASK}, but enable multiple complex-valued phase factors, resulting in \ac{APSK} constellations. These phase bits are mapped onto a set of uniformly spaced phases
\begin{equation}
\textcolor{Set1-E}{
    s_u \in \mathcal{S}_{\text{GPAS}} =  \left\{ \!\exp \! \!\left( \! \j \frac{\pi(2m_{\varphi} \!\!+ \!\!1)}{\tilde{M}_\varphi} \!\!\right) \!\middle|\, m_{\varphi} \!=\! 0, \ldots, \!\tilde{M}_\varphi\!\! -\!\! 1 \!\!\right\}}
\end{equation}
using binary reflected Gray mapping \textcolor{Set1-E}{\mbox{$\beta_{\text{S}}: \{0,1\}^{M_\varphi} \rightarrow \mathcal{S}_{\text{GPAS}}$}}, where \mbox{$\tilde{M}_\varphi = 2^{M_\varphi}$} denotes the number of phase levels. Thus, each $s_u$ is a complex phasor with unit magnitude and quantized angle. The remaining \mbox{$M_{\text{A}} = M - M_\varphi$} bits are used to label the amplitude using binary reflected Gray coding. This generalization transforms the conventional grid-like constellation into a circularly symmetric constellation, resembling \ac{APSK}. Fig.~\ref{fig:GPAS:Constellation_Examples} shows how different combinations of amplitude bits $M_{\text{A}}$ and phase bits $M_\varphi$, with a total of \mbox{$M = 6$} bits, influence the generalized \ac{PAS} constellations and their corresponding \ac{SC} trade-off for a communication \ac{SNR} of \SI{10}{dB}. The configuration with \mbox{$M_{\text{A}} = 2$} amplitude bits and \mbox{$M_\varphi = 4$} phase bits achieves the highest \ac{GMI} across all considered kurtosis values and therefore provides the best \ac{SC} trade-off among the evaluated combinations of amplitude bits $M_{\text{A}}$ and phase bits $M_\varphi$. For this reason, it is selected for comparison in the following analysis. Unlike conventional \ac{PAS}, generalized \ac{PAS} directly produces complex-valued symbols.

Despite this change, the amplitude-phase factorization \mbox{$x_u = a_u \cdot s_u$} still holds, where the amplitude \mbox{$a_u \in \{1, 2, \ldots, 2^{M_\text{A}}\}$} can be shaped independently, while the phasor $s_u$ is uniformly distributed on a unit circle. This allows for assigning the approximately uniformly distributed parity-check bits to the phase. As a result, the encoding structure from conventional \ac{PAS} can be reused, requiring only minor modifications such as allocating more bits to the phase and adapting the inverse sign mapping accordingly.  

\subsubsection{Optimizing Generalized PAS-compatible Constellations using an AE framework}
Similarly to conventional \ac{PAS}, constellations for generalized \ac{PAS} can be efficiently optimized using our \ac{AE} framework by including the structural constraints of generalized \ac{PAS} into the \ac{AE}. The probability of occurrence of the constellation points is the outer product of the trainable target amplitude distribution and a fixed uniform distribution over the phase
\vspace{-0.15cm}
\begin{equation}
    \vect{p}_{\text{GPAS}} = \frac{1}{\tilde{M}_\varphi} \cdot \mathrm{vec} \left( \mathrm{Softmax}\left( \tilde{\vect{p}}_{\RV{a}} \right) \! \otimes \! \vect{1}_{\tilde{M}_\varphi}. \right)\,,
    \vspace{-0.25cm}
\end{equation}
where $\vect{1}_{\tilde{M}_\varphi}$ is an all-one vector of length $\tilde{M}_\varphi$. The probability mass vector $\vect{p}_{\text{GPAS}}$ satisfies the constraints of a conventional \ac{PAS} constellation and replaces the probabilities $\vect{p}$ in the \ac{AE} shown in Fig.~\ref{fig:AE:setup}.

\subsubsection{Low-complexity LLR Calculation}
One key advantage of conventional \ac{PAS} is the low-complexity \ac{LLR} computation enabled by one-dimensional \acp{LUT}. In generalized \ac{PAS}, while the constellation points and their probabilities factor into amplitude and angular components, this factorization does not hold for the channel transition probabilities. As a result, the \ac{LLR} computation is not inherently separable. To address this, we analyze the structure of the exact 2D \acp{LLR}, and identify the feature of the received value that primarily determines the \ac{LLR} values, such as amplitude, phase, or the real or imaginary part. Then, we approximate each \ac{LLR} using only the dominant feature, reducing the computation to a one-dimensional \ac{LUT}. Figure~\ref{fig:PAS:GLLR_values_2D} shows the exact \acp{LLR} for a generalized \ac{PAS} constellation with \mbox{$M_{\text{A}} =2$} amplitude bits (bit 1 \& bit 2) and \mbox{$M_\varphi = 4$} phase bits (bit 3 - bit 6).

For convenience, we recall the \ac{LLR} of bit position $m$, see~\ref{eq:system_model:LLR}
\begin{equation}
    l_{n,m}(y_{\text{c},n}) \! = \!  \log \frac{\sum_{x_{\!n} \in \mathcal{X}^{(0)}_m}{ \! f_{\RV{y}_{\text{c}}|\RV{x}}(y_{\text{c},n}|x_n)}P_{\RV{x}}(x_n)}{\sum_{x_n \in \mathcal{X}^{(1)}_m}{ \! f_{\RV{y}_{\text{c}}|\RV{x}}(y_{\text{c},n}|x_n)}P_{\RV{x}}(x_n)},
\end{equation}
which indicates whether bit $m$ is more likely to be $0$ or $1$ and quantifies the reliability of this decision. We observe that the \acp{LLR} of the first and second bit depend mainly on the amplitude (and not the phase) of the received symbol. Therefore, we transform these \acp{LLR} $l_{1}(\Re\{y_{\text{c}}\},\Im\{y_{\text{c}}\})$ and $l_{2}(\Re\{y_{\text{c}}\},\Im\{y_{\text{c}}\})$ from Cartesian into polar coordinates, where $R = |y_{\text{c}}|$ and $\theta = \sphericalangle \{y_{\text{c}}\}$. Then, we average over the angular dimension $\theta$
\vspace{-0.2cm}
\begin{equation}
    \bar{l}_{m}(R) \approx \frac{1}{2 \pi} \int_0^{2\pi} l_{m}(R, \theta) \, \mathrm{d} \theta, \quad m \in \{1,2\},
    \vspace{-0.1cm}
\end{equation}
resulting in \acp{LLR} that depend solely on the received magnitude $R$. Bits 3 and 4 correspond to the sign of the real and imaginary components, respectively. Thus, we can observe in Fig.~\ref{fig:PAS:GLLR_values_2D} that their \acp{LLR} are primarily influenced by $\Re\{y_{\text{c}}\}$ and $\Im\{y_{\text{c}}\}$, respectively. Therefore, we average over the orthogonal component
\vspace{-0.2cm}
\begin{align}
    \bar{l}_3(\Re \left\{ y_{\text{c}} \right\}) & = \frac{1}{2g} \int_{-g}^{g}  l_3( \Re \{ y_{\text{c}} \} + \mathrm{j}\cdot y) \, \mathrm{d} y \,, \\
    \bar{l}_4(\Im \left\{ y_{\text{c}} \right\}) & = \frac{1}{2g} \int_{-g}^{g}  l_4(x + \mathrm{j} \cdot \Im\{y_{\text{c}}\}) \, \mathrm{d} x \, ,
\end{align}
where $g$ is set to \num{2} covering most received values $y_{\text{c}}$.
Bits 5 and 6 encode further phase information, and we can observe in polar coordinates an oscillatory behavior over the angle $\theta$, with their \acp{LLR} also increasing with magnitude $R$. We propose to approximate the 2D \ac{LLR} functions in polar coordinates $l_5(R, \theta)$ and $l_6(R, \theta)$ each as a product of two one-dimensional functions
\vspace{-0.2cm}
\begin{equation}
    l_{m}(R, \theta) \approx \bar{l}_{m,R}(R) \cdot \bar{l}_{m,\theta}(\theta), \quad m \in \{5,6\},
\end{equation}
where we first compute the phase-dependent term by averaging over the radius with \mbox{$g = 2$}
\vspace{-0.2cm}
\begin{equation}
    \bar{l}_{m,\theta}(\theta) \approx \frac{1}{g} \int_{0}^{g} l_{m}(R, \theta) \, \mathrm{d} R, \quad m \in \{5,6\} \, .
\end{equation}
Subsequently, we estimate the radial dependence by averaging
\vspace{-0.2cm}
\begin{equation}
    \bar{l}_{m,R}(R) \approx \frac{1}{2\pi} \int_{0}^{2\pi} \frac{l_{m}(R, \theta)}{\bar{l}_{m,\theta}(\theta)} \, \mathrm{d}\theta, \quad m \in \{5,6\} \, .
\end{equation}
Note that it is crucial to estimate the phase-dependent component first, as the oscillations lead to a zero average when first averaging over the phase.

\begin{figure}[!t]
    \centering
    \input{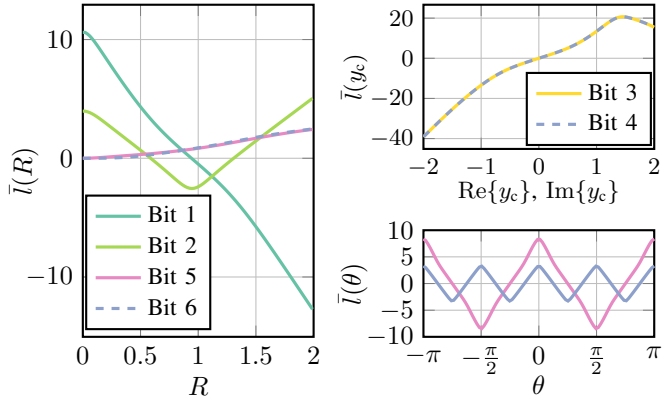}
    \vspace{-0.8cm}
    \caption{One-dimensional \acp{LLR} for the proposed low-complexity \ac{LLR} calculation, obtained by marginalizing the two-dimensional \acp{LLR} shown in Fig.~\ref{fig:PAS:GLLR_values_2D}. The \acp{LLR} are factorized into three domains: (left) amplitude $R$ for bits~1, 2, 5, and 6; (top right) Cartesian components for bits~3 and 4; and (bottom right) phase $\theta$ for bits~5 and 6.}
    \label{fig:PAS:GLLR_values_1D}
    \vspace{-0.45cm}
\end{figure}

Fig.~\ref{fig:PAS:GLLR_values_1D} shows the marginalized \ac{LLR} functions for all six bits, which can now be approximated using one-dimensional \acp{LUT}, enabling low-complexity implementation. In the top-right plot, we observe that bits 3 and 4 depend on their respective Cartesian component in an equivalent manner. This follows from the rotational symmetry of the shaped constellation (see Fig.~\ref{fig:PAS:GLLR_values_2D}) and the fact that they represent coarse phase information, corresponding to the sign of the real and imaginary parts, respectively. Consequently, a single \ac{LUT} suffices for both bits. Furthermore, the radial components of bits 5 and 6 are similar (see left plot) because both bits encode further phase information. Consequently, a single \ac{LUT} suffices for the radial part of both bits. Therefore, only six one-dimensional look-up tables are needed in total: one each for bits 1 and 2, one for bits 3 and 4, one for the radial component shared by bits 5 and 6, and one for their respective phase components. We note that the one-dimensional \acp{LLR} can be approximated using piecewise linear functions, further reducing complexity. However, the specific implementation details are beyond the scope of this paper. Nevertheless, assuming $V$ entries per dimension per \ac{LUT}, this reduces the total storage requirement from $\mathcal{O}(6V^2)$ to $\mathcal{O}(7V)$, offering a significant memory saving. We remark that the proposed approximation generalizes to larger constellations: Amplitude bits remain largely insensitive to phase, the first two phase bits correspond to real and imaginary half-planes, and higher-order phase bits exhibit oscillatory angular patterns. %

\vspace{-0.3cm}
\section{Simulation Results}
\vspace{-0.1cm}
\label{sec:simulation}
In this section, we first validate our derivation of the constellation-dependent detection probability through simulations. Furthermore, we evaluate the \ac{SC} performance and resulting trade-offs of the five constellation shaping methods under both \ac{SMD} and \ac{BMD}, and compare the results to lower and upper bounds on the maximum \ac{MI}. Throughout this section, we optimize constellations with $M = \SI{6}{bit \per symbol}$ for each kurtosis constraint $\tilde{\kappa} \in [1,2]$ independently, under a communications \ac{SNR} of $\mathrm{SNR}_\text{c} = \SI{10}{dB}$ and an increasing penalty factor $d$. The constellations are initialized as \ac{QAM} and optimized by minimizing the loss function~(\ref{eq:AE:overall_loss}) using the Adam optimizer. During training, we increase the batch size from 500 to 10,000, while we decrease the learning rate, with the initial value depending on the shaping method. As discussed in Sec.~\ref{sec:ch2:system_model}, \mbox{$\tilde{\kappa} = 1$} should maximize sensing performance. On the contrary, a larger kurtosis constraint $\tilde{\kappa}$ is expected to improve communications performance, which is maximized for a circular symmetric Gaussian, that has a kurtosis of \mbox{$\kappa = \num{2}$}. A kurtosis constraint $\tilde{\kappa}$ between these two extremes should yield a trade-off between \ac{SC} performance.

\begin{figure}[!t]
    \vspace{0.87cm}
    \input{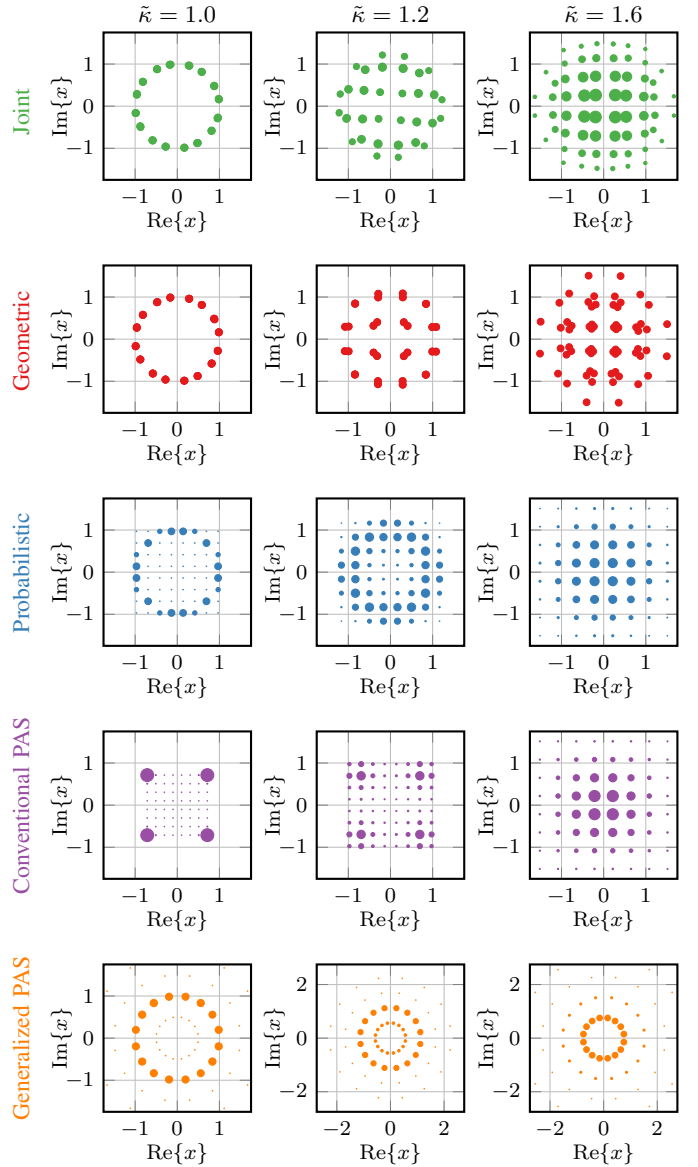}
    \vspace{-0.8cm}
    \caption{Optimized constellations for geometric, probabilistic, and joint constellation shaping as well as conventional and generalized \ac{PAS} under various kurtosis constraints $\tilde{\kappa}$. The size of each constellation point is proportional to its probability and each constellation point has an associated bit label, which is omitted for clarity.}
    \label{fig:res:constellations}
    \vspace{-0.5cm}
\end{figure}

\vspace{-0.2cm}
\subsection{Optimized Constellations}
Fig.~\ref{fig:res:constellations} shows the optimized constellations for \ac{BMD} using geometric, probabilistic (generic, conventional \& generalized \ac{PAS}), and joint constellation shaping for three kurtosis constraints \mbox{$\tilde{\kappa} \in \{1.0,1.2,1.6\}$}. We note that the strongest sensing constraint \mbox{$\tilde{\kappa} = \num{1.0}$} results in unit modulus constellations to reduce the sensing loss term~(\ref{eq:AE:sensing_loss}). In this case, both geometrically and jointly shaped constellations resemble a \ac{PSK} with overlapping constellation points. For generic probabilistic constellation shaping, a \ac{PSK} is only approximated because the constellation points that have the same power do not have equal distances. For a looser sensing constraint \mbox{$\tilde{\kappa} = 1.6$}, the constellations approximate a Gaussian \ac{PDF} to maximize communications performance. In between (\mbox{$\tilde{\kappa} = \num{1.2}$}), a balance between Gaussian and unit modulus \ac{PDF} is learned. Interestingly, the jointly shaped constellations exhibit more geometric shaping characteristics for a small kurtosis constraint $\tilde{\kappa}$ and more probabilistic behavior as the kurtosis constraint $\tilde{\kappa}$ increases.

On the one hand, for conventional \ac{PAS}, the obtained constellation resembles a Gaussian for a weak sensing constraint (large $\tilde{\kappa}$), and is similar to (unconstrained) probabilistic constellation shaping, indicating that the structural constraints imposed by \ac{PAS} do not significantly constrain optimization. However, for a strong kurtosis constraint, probabilistic constellation shaping yields a (blurry) ring, whereas \ac{PAS} resembles a \ac{QPSK}, clearly diverging from the expected circle-like \ac{PDF}. This mismatch arises due to the independent shaping of the real and imaginary components, where the joint \ac{PDF} is the Cartesian product of two one-dimensional marginals and Cartesian structure of conventional \ac{PAS} constrains optimization. On the other hand, generalized \ac{PAS} with \mbox{$M_{\text{A}} = 2$} and \mbox{$M_{\varphi} = 4$} is able to approximate the expected \ac{PSK}-like constellation for a strong sensing constraint.

\begin{figure}[!t]
    \vspace{-0.205cm}
    \centering
    \input{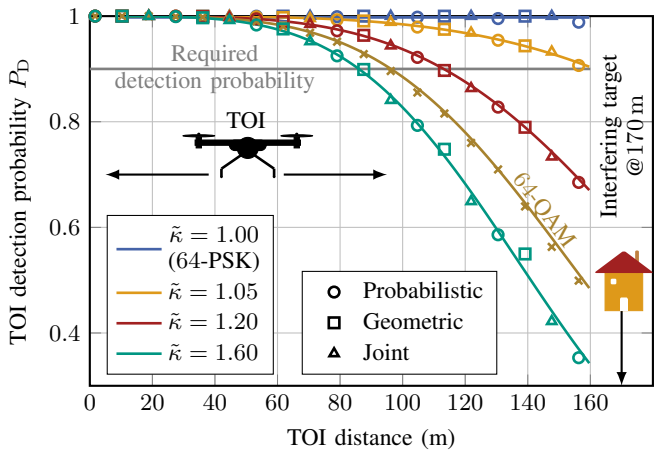}
    \vspace{-0.7cm}
    \caption{Derived detection probability and simulated detection rates of a \ac{TOI}, e.g., a drone, with a radar cross section (\ac{RCS}) of \mbox{$\sigma_{\mathrm{RCS}} = \SI{0.1}{\metre\squared}$} following a Swerling-1 model in the presence of an interfering target, e.g., a building, at \SI{170}{\metre} with an \ac{RCS} of \mbox{$\sigma_{\mathrm{RCS}} = \SI{500} {\metre\squared}$} following a Swerling-0 model~\cite{richards_fundamentals_2014}.}
    \label{fig:res:sensing_performance}
    \vspace{-0.4cm}
\end{figure}

\vspace{-0.2cm}
\subsection{Sensing Performance}
In Fig.~\ref{fig:res:sensing_performance}, we consider a scenario with two targets: a distant interfering target and a nearby \ac{TOI}, which should be detected and whose distance is varied. The \ac{RCS} of the \ac{TOI} follows a Swerling-1 model, whereas the interfering target follows a non-fluctuating Swerling-0 model. Multi-target scenarios are of particular interest because the average \ac{SINR}~(\ref{eq:system_model:mean_SNR_TOI}) depends on the kurtosis and the power reflected by all targets. To verify the impact of the constellation on the detection performance, we evaluate the resulting detection rate for each of the optimized transmit constellations. The simulation setup follows~\cite{braun_ofdm_2014}, with the simulation parameters being the FR2 case from~\cite{mandelli_survey_2023} with \mbox{$N = \num{12672}$} sub-carriers. For simplicity, we use an \ac{FFT} size of \num{12672}, and only one \ac{OFDM} symbol. For \ac{CA}-\ac{CFAR}, we assume a false alarm rate of \mbox{$\PFA = 10^{-3}$} and a sliding window length of $N_{\text{win}}=\num{100}$. To obtain the detection probability, we perform a Monte-Carlo simulation with \num{100} independent Swerling-1 channel realizations. This inherently averages over the \ac{RCS} fluctuations of the \ac{TOI}, such that the mean target power converges to that determined by the mean \ac{RCS}, see~(\ref{eq:system_model:mean_SNR_TOI}). Hence, the impact of the Swerling-1 fluctuations vanishes, and the detection probability is governed by the noise-and-interference power, which depends on the kurtosis~$\kappa$ of the constellation, as described in Sec.~\ref{sec:system_model:detection_probability}.

\begin{figure}[!t]
    \centering
	\input{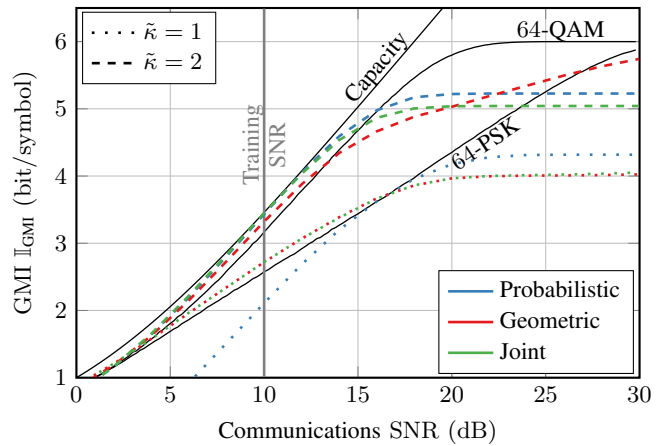}
    \vspace{-0.7cm}
    \caption{Simulated communications performance (\ac{AIR}) of the optimized constellations in comparison to legacy constellation formats.}
    \label{fig:res:comm}
    \vspace{-0.3cm}
\end{figure}

We found that the simulated detection rates (markers) align well with the analytical~(\ref{eq:system_model:detection_prob}) constellation-dependent detection probabilities (curves). This verifies our derivation in Sec.~\ref{sec:ch2:system_model} and shows the impact of the kurtosis $\kappa$ of the constellation on the detection probability. The minor discrepancies stem from the fact that the kurtosis $\kappa$ of the optimized constellations slightly differs from the kurtosis constraint $\tilde{\kappa}$. As expected from (\ref{eq:system_model:detection_prob}) and (\ref{eq:system_model:mean_SNR_TOI}), the detection probability decreases with increasing kurtosis $\kappa$ and depends only on the kurtosis of the constellation, irrespective of the specific shaping method. Moreover, our simulations demonstrate that constellation shaping enables a dynamic adjustment of the detection range. For example, for a required target detection probability of $P_{\mathrm{D}} = \num{0.9}$, the detection range can be varied from $\SI{90}{\metre}$ to beyond $\SI{160}{\metre}$ by modifying the kurtosis $\kappa$ of the constellation.

\vspace{-0.2cm}
\subsection{Communications Performance}
Fig.~\ref{fig:res:comm} shows the \ac{GMI}~\cite[Eq.~14]{Ivanov_BICM} as a function of the communications \ac{SNR} for various sensing constraints $\tilde{\kappa}$. For \mbox{$\tilde{\kappa} = 2$}, all shaping methods reduce the gap to capacity and outperform conventional \num{64}-\ac{QAM} across an \ac{SNR} range of $\SI{10}{dB}$. We note that probabilistic and joint shaping reduce the entropy of the transmit symbols at the design \ac{SNR} of \mbox{$\mathrm{SNR}_\text{c} = \SI{10}{dB}$}, resulting in a saturation of the \ac{GMI} below $\SI{6}{bit/symbol}$ for large \ac{SNR} values. For \mbox{$\tilde{\kappa} = 1$}, the \ac{GMI} of the shaped constellations is similar to that of the \ac{PSK}, although probabilistic constellation shaping performs slightly worse due to unequal distances between the constellation points and the absence of Gray coding. Both geometric and joint constellations slightly outperform the \num{64}-\ac{PSK} because their optimized constellations collapse into several tightly spaced point groups (see Fig.~\ref{fig:res:constellations}), effectively resembling a \num{16}-\ac{PSK}, which achieves a higher \ac{GMI} at low \ac{SNR}. However, within the \ac{SNR} range of Fig.~\ref{fig:res:comm}, these clusters are not distinguishable, practically reducing the number of reliably resolvable points and causing the \ac{GMI} to follow that of a lower-order constellation. In principle, however, these points remain distinguishable as \mbox{$\mathrm{SNR}\to\infty$}. For the geometric constellation with a large kurtosis constraint, the points do not collapse but form locally dense groups (see Fig.~\ref{fig:res:constellations}). These constellation points are not distinguishable at low and moderate \ac{SNR} values, but once the \ac{SNR} exceeds approximately \SI{20}{dB}, even these tightly spaced points become reliably detectable. Consequently, geometric shaping continues to increase its \ac{GMI} toward the maximal achievable entropy of $\log_2(\tilde{M})$ and eventually surpasses probabilistic and joint shaping. The \ac{GMI} of the latter converges to the reduced input entropy due to the non-uniform symbol probabilities in probabilistic shaping, which is optimized for a communications $\mathrm{SNR}_\text{c}$ of $\SI{10}{dB}$. This explains the eventual dominance of geometric shaping for large \ac{SNR}. Note that, in practice, communication systems do not operate over such a wide \ac{SNR} range with a single constellation format. Instead, the constellation format is selected depending on the operating \ac{SNR}.

We remark that the gap between the performance of \ac{PSK}-like constellations with \mbox{$\kappa = 1$} and the \ac{AWGN} capacity reflects the \ac{SC} trade-off. Unit modulus, potentially low-order, constellations achieve optimal sensing performance with only a small gap to \ac{AWGN} capacity at low \ac{SNR}. At higher \ac{SNR}, higher-order modulations are required to increase spectral efficiency for communications, resulting in an increasing \ac{SC} trade-off. Yet, the \ac{SNR} in wireless systems is typically limited by power constraints and multipath fading. As a result, many practical systems operate in moderate \ac{SNR} regimes, where \num{64}-\ac{QAM} is commonly used, making our parametrization particularly relevant for real-world deployments.

\begin{figure}[!t]
    \centering
    \input{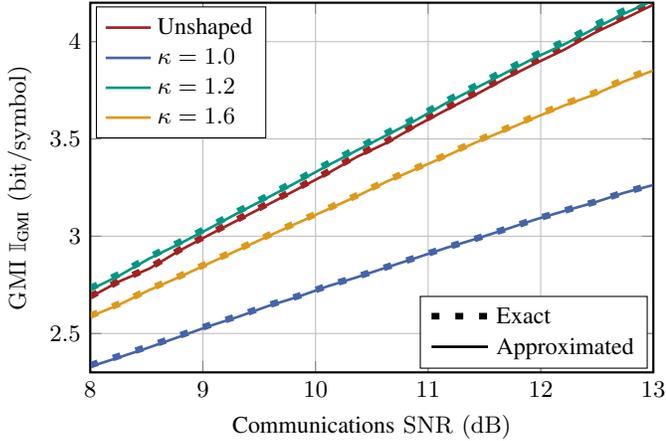}
    \vspace{-0.8cm}
    \caption{Communication performance comparison between exact and low-complexity approximate \ac{LLR} computation for unshaped and shaped generalized \ac{PAS} constellations.}
    \label{fig:GPAS:comm_performance}
    \vspace{-0.5cm}
\end{figure}

Next, we assess our proposed low-complexity \ac{LLR} calculation for generalized \ac{PAS}, and compare it against the exact \ac{LLR} calculation~(\ref{eq:system_model:LLR}). Fig.~\ref{fig:GPAS:comm_performance} shows the \ac{GMI} as a function of the communications \ac{SNR} for both unshaped (equiprobable transmit symbols) and shaped generalized \ac{PAS} constellations, respectively. It can be observed that the curves corresponding to the exact and approximated \ac{LLR} calculations match closely with a maximum deviation of less than \SI{0.016}{bit/symbol}. This confirms that the proposed approximation enables low-complexity \ac{LLR} computation with negligible performance loss.

\begin{figure}[!t]
\centering
    \input{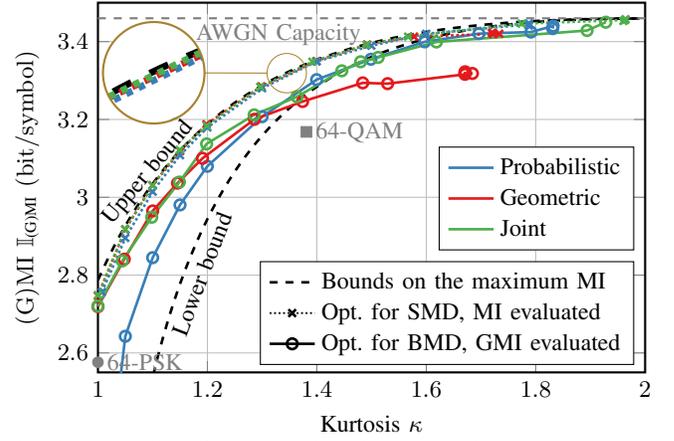}
    \vspace{-0.7cm}
    \caption{Comparison of the derived bounds on the maximum \ac{MI} and the \ac{SC} trade-off in terms of \ac{GMI} and \ac{MI} for different shaping methods and legacy modulation formats. Markers represent individually optimized constellations. For large kurtosis constraints, the kurtosis of the optimized constellations saturates, indicating that the maximum \ac{AIR} is achieved with a kurtosis lower than the constraint. The kurtosis saturates below \num{2} for all shaping methods.}
    \label{fig:results:random_determinisitic}
    \vspace{-0.5cm}
\end{figure}

\vspace{-0.25cm}
\subsection{Sensing \& Communications Trade-off}
\vspace{-0.05cm}
Finally, we demonstrate how constellation shaping can govern the \ac{DRT} in \ac{ISAC} systems by analyzing the \ac{MI} and \ac{GMI} as a function of the kurtosis $\kappa$ for geometric, probabilistic, and joint shaping. These results are compared in Fig.~\ref{fig:results:random_determinisitic} to conventional modulation formats as well as the lower and upper bounds on the maximum \ac{MI} derived in Sec.~\ref{sec:bound} to assess both the derived bounds and the effectiveness of our optimized constellations.

First, we observe that the \ac{AIR} increases monotonically with increasing kurtosis across all shaping methods and for both bounds. It exhibits a particularly steep rise for lower kurtosis values and approaches the \ac{AWGN} capacity for a kurtosis of \mbox{$\kappa = 2$}. Notably, the gap between the upper and lower bounds on the maximum \ac{MI} narrows with increasing kurtosis, dropping below \SI{0.1}{bit/symbol} for \mbox{$\kappa \geqslant 1.35$} and vanishing at \mbox{$\kappa = 2$}. This is because for $\kappa = 2$, the involved \acp{PDF} become Gaussian and the \ac{EPI} is tight. With decreasing kurtosis constraint, the deviation of the entropy-maximizing transmit \ac{PDF} from a Gaussian increases, making the \ac{EPI} and, therefore, the lower bound increasingly loose.

\begin{figure}[!t]
\centering
    \input{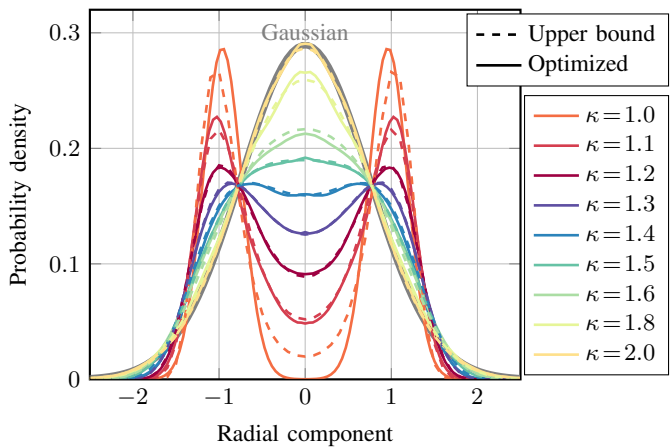}
    \vspace{-0.8cm}
    \caption{Comparison of the empirical received signal \ac{PDF} resulting from the optimized constellations and the theoretical maximum-entropy \ac{PDF} used to derive the upper bound for various kurtosis constraints, along with a Gaussian as reference. Note that the original \acp{PDF} are two-dimensional. The displayed \acp{PDF} are obtained by averaging one-dimensional marginals computed along \num{64} uniformly spaced radial directions in the range $[0, \pi]$.}
    \label{fig:results:distribution_received_signal}
    \vspace{-0.4cm}
\end{figure}

Next, we observe that the constellations optimized for \ac{SMD} yield a \ac{MI} that closely approaches the upper bound~(\ref{eq:bound:final_bound}). In Fig.~\ref{fig:results:distribution_received_signal}, we compare the empirical \ac{PDF} of the received signal resulting from the optimized constellations to the maximum-entropy \ac{PDF} used to derive the upper bound~(\ref{eq:Bound:Ansatz_Maximum_entropy}). For $\kappa = 2$, both \acp{PDF} converge to the circularly symmetric Gaussian \ac{PDF}, which achieves the capacity for an \ac{AWGN} channel. At lower kurtosis values, the two \acp{PDF} diverge slightly, as the upper bound does not constrain the output to be realizable under the system model, resulting in a gap between the achieved \ac{MI} and the derived bound~(\ref{eq:bound:final_bound}). Nevertheless, the gap in Fig.~\ref{fig:results:random_determinisitic} is less than \SI{0.04}{bit/symbol} at \mbox{$\kappa = 1$} and below \SI{0.01}{bit/symbol} for \mbox{$\kappa \geqslant 1.05$}. This demonstrates the effectiveness of the \ac{AE}-based constellation optimization and shows that the upper bound can be closely approached by constellation shaping.

Now, we focus on the \ac{GMI} as an \ac{AIR} in practical communication systems using \ac{BMD}. As expected, the \ac{GMI} is consistently lower than the \ac{MI}, with the performance gap depending on both the kurtosis constraint and the shaping method. 

We observe that probabilistic constellation shaping effectively approaches the capacity and outperforms geometric constellation shaping if the sensing constraint is loose (large $\tilde{\kappa}$). On the contrary, geometric constellation shaping outperforms probabilistic constellation shaping for strict sensing constraints \mbox{$\tilde{\kappa} < 1.3$}. While this is also true for the \ac{MI}, the performance differences are far more pronounced in terms of \ac{GMI}. This highlights the importance of directly optimizing the \ac{GMI}, rather than the \ac{MI} as done in prior work~\cite{du_probabilistic_2023,du_reshaping_2023,liu_probabilistic_2025}.

Unlike \ac{MI}, which measures symbol-wise reliability, \ac{GMI} reflects bit-level reliability and is therefore sensitive to the binary labeling of the constellation points. However, for geometrically shaped constellations a Gray labeling may not exist. Gaussian-like constellations at high kurtosis values lack the regular grid structure needed for Gray coding, resulting in a performance gap at high kurtosis values. In probabilistic shaping, the bits are correlated, but \ac{BMD} computes the \acp{LLR} independently, see~(\ref{eq:system_model:LLR}), (\ref{eq:system_model:GMI}), leading to an \ac{AIR} loss, especially under strong shaping, i.e., at low kurtosis. Moreover, the resulting constellations are non-uniformly spaced and lack Gray coding between diagonally placed symbols (see Fig.~\ref{fig:res:constellations}), which further degrades the \ac{GMI} for low kurtosis values. To validate that the observed gap is indeed a \ac{BMD} artefact rather than a result of the optimization process, we computed the \ac{MI} of the constellations optimized for \ac{BMD} and found it closely approaches the \ac{MI} of the constellations optimized for \ac{SMD}, confirming that the degradation stems from \ac{BMD}.

\textbf{Remark:} This makes geometric and probabilistic constellation shaping well-suited for applications where sensing or communications performance is prioritized, respectively. Joint shaping consistently exhibits strong performance across all regimes, effectively leveraging the advantages of both geometric and probabilistic shaping. It performs similarly to geometric and probabilistic constellation shaping at low and high kurtosis values, respectively. Compared to conventional \num{64}-\ac{QAM}, joint constellation shaping achieves a \num{0.16} bit/symbol higher \ac{GMI} at the same kurtosis~$\kappa$. This means that joint shaping increases the communication throughput while maintaining the same detection probability for sensing. Conversely, joint shaping can reduce the kurtosis~$\kappa$ by \num{0.19} while achieving the same \ac{GMI}, which directly translates into a higher detection probability of the \ac{TOI} at equal communication performance. In addition, while \num{64}-\ac{PSK} and \num{64}-\ac{QAM} offer only two discrete operating points, joint shaping enables a continuous adjustment of the \ac{AIR}-kurtosis operating point. This allows for a flexible and fine-grained trade-off between communication throughput and sensing detection probability. Furthermore, remember that the self-interference from the limited isolation between the transmitter and receiver paths manifests as a target with a large virtual \ac{RCS}, and that the noise-and-interference~(\ref{eq:system_model:noise_and_interference_power}) floor scales with the product of the kurtosis and the summed power of all targets. Consequently, if the self-interference is stronger than the received reflections of the targets, it dominates the noise-and-interference floor, which scales with the kurtosis of the employed constellations. Hence, reducing the kurtosis through constellation shaping can reduce the noise-and-interference floor caused by the self-interference and, consequently, increase the detection probability of a \ac{TOI} with small \ac{RCS}.

\begin{figure}[!t]
    \centering
    \input{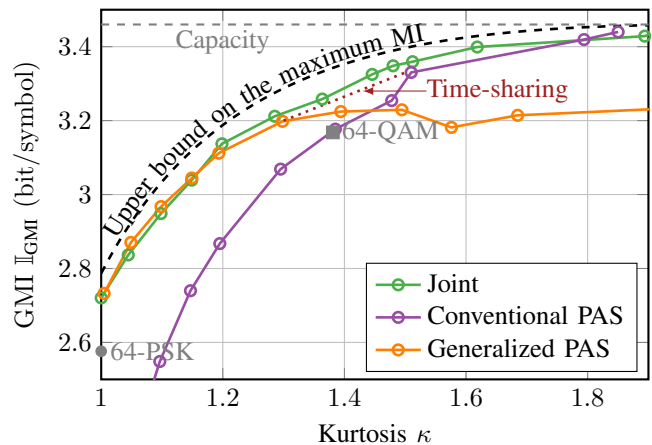}
    \vspace{-0.3cm}
    \caption{Trade-off between communication performance (GMI) and sensing performance (kurtosis) for conventional and generalized \ac{PAS}, compared to joint constellation shaping and conventional modulation formats.}
    \label{fig:GPAS:rand_det_tradeoff}
    \vspace{-0.5cm}
\end{figure}

To quantify the performance gains of low-complexity constellation shaping, Fig.~\ref{fig:GPAS:rand_det_tradeoff} compares conventional and generalized \ac{PAS} in terms of their \ac{SC} trade-off and benchmarks them against joint constellation shaping and standard modulation formats. Conventional \ac{PAS} performs best under weak kurtosis constraints, i.e., high kurtosis, closely approaching the \ac{AWGN} capacity. However, as discussed previously, it suffers from a significant penalty when circular constellations with low kurtosis are required. In this regime, generalized \ac{PAS} matches the performance of joint constellation shaping. In the intermediate regime, a small gap remains between the \ac{PAS} variants and joint shaping, which could, in principle, be closed by time-sharing between both shaping configurations.

\textbf{Remark:} A combination of conventional and generalized \ac{PAS} outperforms legacy modulation formats and achieves nearly the same \ac{SC} performance as joint shaping, which fully exploits all available degrees of freedom. However, in contrast to joint shaping, both conventional and generalized \ac{PAS} maintain low complexity at the transmitter and receiver. This makes them well-suited for 6G \ac{ISAC} systems, where maximizing performance while enabling a flexible trade-off between \ac{SC} under tight implementation constraints will be crucial.

\vspace{-0.4cm}
\section{Conclusion}
\vspace{-0.1cm}
We optimized and systematically compared geometric, probabilistic, and joint constellation shaping for \ac{OFDM}-\ac{ISAC} using a bitwise \ac{AE} framework, maximizing the \ac{MI} and the \ac{GMI} under a sensing target detection probability constraint. We showed analytically and confirmed through simulations that the detection probability depends solely on the kurtosis of the constellation. Next, we derived lower and upper bounds on the maximum \ac{MI} and showed that the optimized constellations achieve \ac{MI} values close to the upper bound.

In practical systems using \ac{BMD}, geometric shaping achieves a higher \ac{GMI} under strict sensing constraints, while probabilistic shaping performs better under relaxed sensing constraints. Notably, our proposed joint shaping approach combines the strengths of both geometric and probabilistic constellation shaping, significantly outperforming legacy constellation formats and offering a flexible \ac{SC} trade-off.

Furthermore, we revisited \ac{PAS} as a promising candidate for a practical implementation of constellation shaping and showed that its independent shaping of in-phase and quadrature components restricts the circular symmetry needed for constellations with low kurtosis. To overcome this, we generalized \ac{PAS} to yield circularly symmetric constellations and proposed a low-complexity \ac{LLR} approximation with negligible performance loss, preserving the original low-complexity advantage. Our results show that combining conventional and generalized \ac{PAS} achieves near-joint shaping \ac{SC} performance with low complexity.

Thus, constellation shaping enables dynamically adjustable, high-performance \ac{SC} at low complexity, making it a compelling solution for flexible and efficient 6G \ac{ISAC} systems.

Future research includes extending the proposed framework to frequency-selective communication channels, where the communications \ac{SNR} varies across subcarriers and power allocation becomes relevant, as discussed in Sec.~\ref{sec:ch2:system_model}. Additional directions include incorporating the Doppler effect, i.e., velocity estimation, for sensing and studying constellation shaping under practical hardware impairments.

\appendices
\vspace{-0.2cm}
\section{}
\label{sec:App_B}
\vspace{-0.05cm}
To show that the kurtosis $\kappa$ of a conventional $\widetilde{M}$-\ac{QAM} approaches \mbox{$\kappa \approx 1.4$} as \mbox{$\widetilde{M}\to\infty$} (see Sec.~\ref{sec:system_model:detection_probability}), we use that, in this limit, the constellation converges to an independently and uniformly distributed random variable on $\mathcal{U}[-\sqrt{3/2},\sqrt{3/2}]$ in both dimensions, i.e.,
\vspace{-0.2cm}
\begin{equation}
    \RV{x} = \RV{x}_{\text{Re}} \!+ \!\j \RV{x}_{\text{Im}}, \!\! \!\quad
    \RV{x}_{\text{Re}},\RV{x}_{\text{Im}} \!\sim\! \mathcal{U}\big[\!-\!\sqrt{3/2},\sqrt{3/2}\big], \!\! \! \quad \RV{x}_{\text{Re}} \!\perp\! \RV{x}_{\text{Im}}.
    \vspace*{-0.2cm}
\end{equation}
The kurtosis then directly follows as
\vspace{-0.15cm}
\begin{align}
    \mathbb{E}\{|\RV{x}|^4\} & = \! \mathbb{E}\!\left\{ \! \left(\RV{x}^2_{\text{Re}} \!+\! \RV{x}^2_{\text{Im}}\right)^{\!2} \!\right\} \!=\! 2 \! \cdot \! \mathbb{E}\!\left\{\RV{x}^4_{\text{Re}}\right\}\! +\! 2 \!\cdot\!  \left(\mathbb{E}\!\left\{\RV{x}^2_{\text{Re}}\right\}\right)^2 \\
    & = 2 \cdot \frac{9}{20} + 2\cdot\left( \frac{1}{2} \right)^2 = \frac{7}{5},
    \vspace*{-0.25cm}
\end{align}
where we used \mbox{$\mathbb{E}\{\RV{s}^4\} = a^4/5$} and \mbox{$\mathbb{E}\{\RV{s}^2\} = a^2/3$} for \mbox{$\RV{s} \sim \mathcal{U}[-a,a]$} with \mbox{$a=\sqrt{3/2}$}.

\vspace{-0.2cm}
\section{}
\label{sec:App_A}
\vspace{-0.05cm}
In this appendix, we show that solving the nonlinear system~(\ref{eq:bound:non_linear_eq_system}) is equivalent to solving~(\ref{eq:bound:gamma2_1D}) and inserting the result into~(\ref{eq:bound:gamma0_1D}) and~(\ref{eq:bound:gamma4_1D}). We begin by rewriting~(\ref{eq:bound:non_linear_eq_system}) for $q = 0, 1, 2$ as
\vspace{-0.2cm}
\begin{align}
    C_q & = \int_0^{2\pi} \int_0^\infty r^{2q+1} e^{\gamma_0 + \gamma_2 r^2 + \gamma_4 r^4} \,\mathrm{d}r\,\mathrm{d}\theta \label{eq:appdx:a} \\
     & = 2\pi \e^{\gamma_0} \int_0^\infty r^{2q+1} e^{\gamma_2 r^2 + \gamma_4 r^4} \,\mathrm{d}r \label{eq:appdx:b} \\
     & = \pi \e^{\gamma_0} \int_0^\infty u^{q} e^{\gamma_2 u + \gamma_4 u^2} \,\mathrm{d}u, \label{eq:appdx:c} \\
     & = \pi \e^{\gamma_0} \int_0^\infty u^{q} \e^{\frac{\gamma^2_2}{4\gamma_4}} \e^{- |\gamma_4| \left( u - \frac{\gamma_2}{2\gamma_4} \right)^2} \,\mathrm{d}u \label{eq:appdx:d} \\
     & = \pi \e^{\gamma_0} \e^{\frac{\gamma^2_2}{4\gamma_4}} \!\!\!\!\!\!\!\!\!\! \int\limits_{-\gamma_2/(2|\gamma_4|)}^\infty \!\!\!\!\!\!\left(v + \frac{\gamma_2}{2|\gamma_4|}\right)^{q}  \e^{- |\gamma_4| v^2}  \,\mathrm{d}v \label{eq:appdx:e}
     \vspace*{-0.1cm}
\end{align}
where we transform a 2D Cartesian complex-plane integral to polar coordinates with \mbox{$Z = r \e^{\j \theta}$} in~(\ref{eq:appdx:a}), and substitute \mbox{$u = r^2$} in~(\ref{eq:appdx:c}). In~(\ref{eq:appdx:d}), we complete the square in the exponent and set \mbox{$\gamma_4 = - |\gamma_4|$}, since the integral diverges for \mbox{$\gamma_4 > 0$}. Finally, we substitute \mbox{$v = u - \frac{\gamma_2}{2\gamma_4}$} in~(\ref{eq:appdx:e}).

Using the integrals~\cite[entries:~3.321.2,~3.321.4,~3.321.5]{gradshteyn2007}, we obtain closed-form expressions
\vspace{-0.2cm}
\begin{align}
    C_0 & = \pi \e^{\gamma_0} \frac{1}{2} \sqrt{\frac{\pi}{|\gamma_4|}} \mathrm{erfc}\!\left(\frac{-\gamma_2}{2\sqrt{|\gamma_4|}}\right) \e^{\frac{\gamma_2^2}{4|\gamma_4|}}, \label{eq:apdx:C0} \\
    C_1 & = \pi \e^{\gamma_0} \left[ \frac{1}{2|\gamma_4|} + \frac{\gamma_2 \sqrt{\pi}}{4|\gamma_4|^{3/2}} \mathrm{erfc}\!\left(\frac{-\gamma_2}{2\sqrt{|\gamma_4|}}\right) \e^{\frac{\gamma_2^2}{4|\gamma_4|}} \right],\label{eq:apdx:C1} \\
    \begin{split}
    C_2 & =  \pi \e^{\gamma_0} \left[ \frac{\gamma_2}{4|\gamma_4|^2}+ \left( \frac{1}{4|\gamma_4|}+\frac{\gamma_2^2}{8|\gamma_4|^{2}}\right) \right. \\& \quad \qquad \quad\cdot \left.\sqrt{\frac{\pi}{|\gamma_4|}} \mathrm{erfc}\!\left(\frac{-\gamma_2}{2\sqrt{|\gamma_4|}}\right) \e^{\frac{\gamma_2^2}{4|\gamma_4|}} \right]. \end{split} \label{eq:apdx:C2}
    \vspace*{-0.15cm}
\end{align}

Using $C_0 = 1$~(Tab.~\ref{tab:bound:constraints}), we substitute the common term in~(\ref{eq:apdx:C0}) into~(\ref{eq:apdx:C1}) and~(\ref{eq:apdx:C2}) yielding
\vspace{-0.15cm}
\begin{align}
    C_1 & = \frac{1}{2|\gamma_4|} \left( \pi \e^{\gamma_0} + \gamma_2\right), \label{eq:apdx:C1_short}\\
    C_2 & = \frac{\gamma_2}{4|\gamma_4|^2} \left( \pi \e^{\gamma_0} + \gamma_2 \right) + \frac{1}{2|\gamma_4|}. \label{eq:apdx:C2_short}
    \vspace{-0.1cm}
\end{align}
Inserting~(\ref{eq:apdx:C1_short}) into~(\ref{eq:apdx:C2_short}) and then back into~(\ref{eq:apdx:C1_short}) yields~(\ref{eq:bound:gamma4_1D}) and~(\ref{eq:bound:gamma0_1D}). Finally, inserting both~(\ref{eq:bound:gamma0_1D}) and~(\ref{eq:bound:gamma4_1D}) in~(\ref{eq:apdx:C0}) yields~(\ref{eq:bound:gamma2_1D}).

\vspace{-0.2cm}

 \bibliography{IEEEabrv,references_copy.bib}
 \bibliographystyle{IEEEtran}

\end{document}